\documentclass[aps,pre,preprint,groupedaddress]{revtex4-1}

\usepackage{amsmath}
\usepackage{amssymb}
\usepackage{graphicx} 
\usepackage{caption}
\usepackage{subcaption}
\usepackage{float}
\usepackage{bm}
\usepackage[usenames,dvipsnames]{xcolor}
\DeclareGraphicsExtensions{.pdf,.png,.jpg}
\graphicspath{{figs/}}

\renewcommand{\Re}{\ensuremath \textnormal{Re}}	
\newcommand{\Cab}{\ensuremath \textnormal{Ca\textsubscript{b}}}
\newcommand{\Cat}{\ensuremath \textnormal{Ca\textsubscript{t}}}
\renewcommand{\vec}[1]{\ensuremath\bm{#1}}

\begin{document}


\title{Non-equilibrium dynamics of initially spherical vesicles in general flows}


\author{Afsoun Rahnama Falavarjani}
\author{David Salac}
\email[]{davidsal@buffalo.edu}
\affiliation{University at Buffalo}

\date{\today}

\begin{abstract}
	Many vesicles have a spherical resting shape and exposure to fluid flows induces an exchange between 
	sub-optical area and visible (systematic) deformation, while the total area is conserved. 
	The dynamics which controls the exchange between sub-optical and visible area depends on membrane properties such as 
	bending rigidity and initial tension. Conversely, observation of these dynamics can be used to determine the membrane properties.
	The goal of this work is to create a general numerical model which accounts for the exchange
	between sub-optical and visible area.
	Unlike prior modeling efforts, the model does not pre-assume a shape type, such as nearly-spherical,
	or applied flow field, allowing the model to capture a wider variety of flow conditions.
	Based on implicit interface tracking and using a volume-preserving multiphase Navier-Stokes
	solver, the model is compared to several experimental results, showing excellent agreement.
	It is used to explore regimes not possible with previously published models, such as the
	variable viscosity case, and how these system properties influence experimentally measurable 
	parameters such as deformation parameter. By creating a more generalized framework for the modeling
	of vesicles with sub-optical area, it will now be possible to make predictions on vesicle material properties
	from a wider variety of experimental results.
\end{abstract}


\maketitle

\section{Introduction}
Giant liposome vesicles are enclosed bag-like membranes 
composed of a lipid bilayer. The lipids themselves are
amphiphilic molecules which have a tendency to form a ``head-out"
arrangement when exposed to polar solvents such as water~\cite{C7RA02746J}
and will form vesicles once a critical patch size is reached~\cite{Huang2910}.
Due to the low solubility of lipid molecules, it is possible to ignore
any exchange of lipid molecules between the bilayer membrane and the surrounding
fluid, leading to a membrane containing a fixed number of molecules~\cite{dimova2019giant}.
When combined with the closed-packed nature of the lipids on the membrane
it is common to assume that the total surface area of a vesicle is conserved.
Typically it is assumed that all of the area is in the visible regime, and this
assumption has been extensively used to model vesicles in 
shear~\cite{Hatakenaka2011,kantsler2005orientation,kraus1996, Yazdani2012, PhysRevE.86.066321, vlahovska2007dynamics,Zabusky2011,zhao2011}
and Poiseuille~\cite{danker2009vesicles} flows, through constrictions~\cite{Barrett2016}, 
under the influence of gravity~\cite{kraus1995}, and 
exposure to electric fields~\cite{kolahdouz2015,schwalbe2011a,HU201666,C5SM00585J,Salipante2014}.

\begin{figure}
	\centering
	\includegraphics[width=8.5cm]{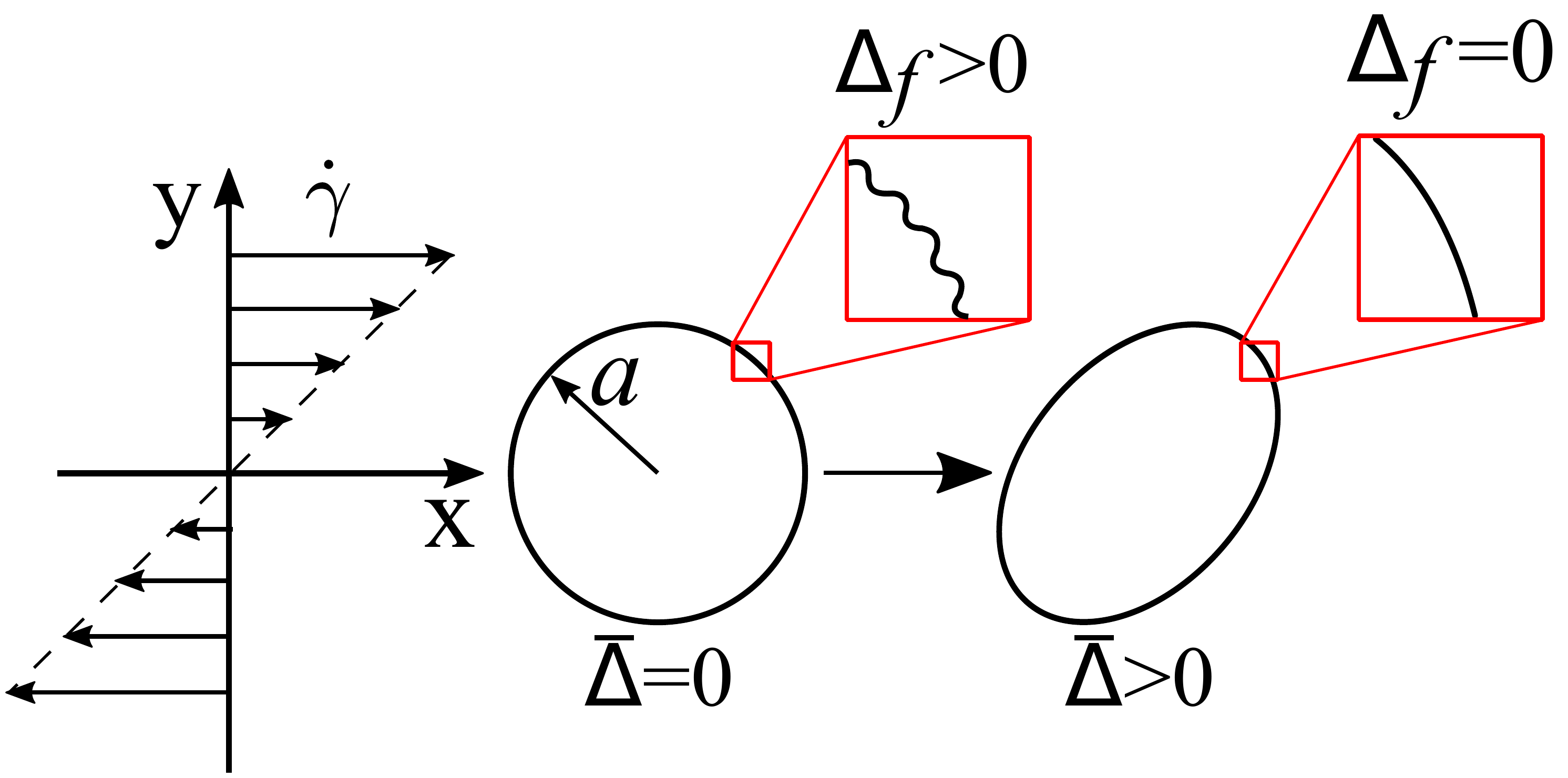}
	\caption{Schematic of the excess area stored in membrane fluctuations and their flattening in shear flow. Initially the vesicle
		has a radius of $a$ with zero systematic (visible) excess area, $\bar{\Delta}$. All excess area is stored in sub-optical area, $\Delta_f>0$. During elongation the sub-optical area is transferred into visible area such that $\bar{\Delta}>0$ and $\Delta_f=0$.}
	\label{fgr:fluctu}
\end{figure}

Despite this constraint on surface area, experiments have demonstrated that the
visible area of vesicles does in fact change under certain situations, such as
the application of electric fields~\cite{Dimova2007,Salipante2014,Yu2015}. It has been hypothesized that
this behavior does not, in fact, run counter to a constant surface area.
Instead, the total surface area is actually composed of two contributions:
the visibly apparent area and an area stored in sub-optical fluctuations
which are on the scale of the membrane thickness, see Fig.~\ref{fgr:fluctu}. 
When an external force is applied to an initially spherical vesicles, 
the area stored in sub-optical fluctuations
are pulled (transferred) into the visible area, thus increasing the apparent area while
the total area is held constant.

The exchange between sub-optical and visible area can be used to determine properties of the bilayer such as bending rigidity,
which provides the energetic penalty associated with the bending of the membrane and is typically on the order
of $10^{-19}$ J~\cite{DIMOVA2014}. 
For example, micropipette aspiration experiments performed by Evans and Rawicz~\cite{Evans1990} determined a relation between tension and visible area expansion. 
They show that in the low tension regime where thermal fluctuations dominate
the deformation of the bilayer, the increase of the tension and the relative area dilation can be used to directly measure bending rigidity. 
Shear flow experiments performed by de Haas et al.~\cite{deHaas1997deformation} used a deformation model to determine the likely membrane bending rigidity of the vesicles explored.
Other experimental techniques which use the increase in area due to the flattening of sub-optical fluctuations include the 
use of optical tweezers~\cite{Brown2011,Delabre2015} and electric fields~\cite{kummrow1991deformation}. Other examples of experimental techniques
for membrane property determination are available in the literature, such as the review by Dimova~\cite{DIMOVA2014}.

There have been many theoretical and numerical investigations of vesicles with zero sub-optical area~\cite{Misbah2006,salac2012reynolds,Kolahdouz2015b,RAHIMIAN2015766,Quaife2019,danker2009vesicles}.
Fewer investigations starting with the assumption of a spherical vesicle and an assumption of sub-optical area exist.
Samples include the use of spherical harmonics to determine material properties from stretching via optical tweezers and 
thermal fluctuations~\cite{Zhou2011} and transient solutions of vesicles in electric fields by assuming a nearly-spherical shape~\cite{Zhang2013}.
Another example is the work by Yu et al~\cite{Yu2015}, which explored the relaxation of a vesicle after the cessation of an external electric field. Using the uniform, entropic tension~\cite{Seifert1997} and assuming that the shape can be described by a single-mode perturbation from an ellipse they were able to obtain a closed form solution for the vesicle shape. This was compared to experimental results to extract the bending rigidity of the vesicles. While this work was able to capture material properties, it is limited to only vesicles relaxing back to a spherical shape and in the absence of any externally driven flow.

In this work a general numerical framework to model liposome vesicles~\cite{salac2011, GERA2018, kolahdouz2015} is extended to account for the exchange between sub-optical area and visible area. Unlike many prior modeling efforts, we do not make \textit{a priori} assumptions about the vesicle shape (such as nearly-spherical) or the underlying flow field, allowing for a wider variety of flow conditions to be investigated. This could, for example, allow for the design of experiments which provide the greatest amount of useful information to determine membrane material properties or determine said properties from already-completed experiments.

After describing the general mathematical and numerical framework, the model will be directly compared to a number of different experimental results. The results will also demonstrate under which conditions other models no longer hold.
While this method assumes that a uniform tension exists as a function of the increase in visible area, 
which was developed for quasi-spherical vesicles described using spherical harmonics~\cite{Seifert1999}, the numerical results match qualitatively well with published experimental works for small-to-moderate visible excess areas.

\section{Mathematical Model and Numerical Methods}
Here it is assumed that the vesicles are made of a closed fluid phase lipid bilayer encapsulating a fluid which can differ from the fluid outside of the vesicle. The particular shape of the vesicle and the fluid field in the
vicinity of the membrane depends not only on any externally applied flows but also the forces induced by the membrane itself. 
In particular the effect of membrane bending and area-dependent tension need to be accounted for. Additionally, a numerical method
which conserves the enclosed volume and corrects for any volume losses due to numerical errors must be created. In this section a description
of the isotropic tension, the multiphase Navier-Stokes equations with singular membrane force contributions, and the numerical methods used 
are described.

\subsection{Isotropic tension}

Vesicles are assumed to have constant enclosed volume due to membrane impermeability and constant total area.
A common measure of a vesicle is the excess area, which measures the difference between the area of the vesicle to that of a sphere with the same volume. 
Given a three-dimensional vesicle with volume $V$ and area $A$, the deviation of the vesicle shape from a sphere can be quantified by the 
excess area $\Delta = A/a^2-4\pi$, where $a$ is the radius of a vesicle with the same volume and is given by $a=\big(3V/4\pi)^{(1/3)}$. 
An alternative measure, not used in this work, is the reduced volume, which is a non-dimensional quantity indicating the fraction of volume contained by the vesicle 
compared to the volume of a spherical vesicle with the same surface area. 
The reduced volume of a vesicle is given by $\upsilon=3V/(4\pi R_0^3)$, where $R_0=(A/4\pi)^{(1/2)}$.
For a sphere, the excess area is $\Delta=0$ while the reduced volume is $\upsilon=1$. For a vesicle where sub-optical fluctuations have been converted to visible area
we have $\Delta>0$ and $\upsilon<1$.

Following the prior discussion, it is common to assume that the total excess area of a vesicle, $\Delta$, is fixed. Let the visible (systematic) excess area be given by $\bar{\Delta}$. 
Any excess area in sub-optical fluctuations is given by $\Delta_f$ such that 
$\Delta_f = \Delta-\bar{\Delta}$. 
When a vesicle is exposed to an external flow, the gradual flattening of the undulations of the membrane transfers the excess area from the fluctuations, $\Delta_f$, to systematic deformation, $\bar{\Delta}$, which leads to an
increase in apparent area, $\Delta A = A-A_0$, where $A_0$ is the area of the vesicle in the absence of any external forces or flow.
This increase in area gives rise to an isotropic tension, $\sigma$~\cite{Seifert1997}. 
At low tensions (small values of $\Delta A/A_0$), called the entropic regime, the tension is related to the change in apparent area through~\cite{Evans1990,Borghi_2003}
\begin{equation}\label{eqn:IsotropicTen}
	\dfrac{\Delta A}{A_0} = \frac{K_B T}{8\pi\kappa_c}\ln\left(\dfrac{1+A\sigma/(24\pi\kappa_c)}{1+A\sigma_0/(24\pi\kappa_c)}\right)
\end{equation}
where $\sigma_{0}$ is the surface tension of an unforced vesicle, $\kappa_c$ is the bending rigidity of the membrane, while $K_B$ and $T$ are the Boltzmann constant and absolute temperature, respectively.
After all sub-optical fluctuations are flattened, the membrane might undergo slight stretching. In this case the tension increases linearly with the change in area:
\begin{equation}
	\dfrac{\Delta A}{A_0} \approx \frac{\sigma-\sigma_0}{K_a},
\end{equation}
where $K_a$ is the stretching modulus of the membrane.
The cross-over point from entropic (exponential) to linear tension is typically given by~\cite{dimova2002hyperviscous}
\begin{equation}\label{eqn:CrossTen}
	\sigma_c = \frac{K_B T K_a}{8\pi \kappa_c}.
\end{equation}
For typical values of bending and stretching moduli $\kappa_c\approx 20K_B T$ and $K_a\approx 0.2$ N/m 
the cross-over tension is estimated to be on the order of $\sigma_c \approx 4\times 10^{-4}$ N/m, which is greater than typical values obtained in our simulations. Therefore,
it is not expected that any stretching of the membrane will occur, justifying the constant total surface area assumption.

The complete range of tension can thus be described by~\cite{Evans1990,Borghi_2003}
\begin{equation}\label{eqn:CompleteTen}
	\dfrac{\Delta A}{A_0} = \dfrac{K_B T}{8\pi\kappa_c}\ln\left(\dfrac{1+A\sigma/(24\pi\kappa_c)}{1+A\sigma_0/(24\pi\kappa_c)}\right) + \dfrac{\sigma-\sigma_0}{K_a}.
\end{equation}
In the low-tension regime, Eq.~\eqref{eqn:CompleteTen} is dominated by the logarithmic term and after crossover to the high-tension regime the elastic stretching term becomes important. 
Note that for planar membranes the $24\pi$ terms in Eqs.~\eqref{eqn:IsotropicTen} and \eqref{eqn:CompleteTen} are 
replaced with $\pi^2$~\cite{shi2014dynamics}.
In our investigations the difference between the two terms is less than 0.02 percent. 
Additionally, while the form used in Eq.~\eqref{eqn:CompleteTen} is used in this work, it is common
to assume that $A\sigma_0\gg24\pi\kappa_c$ and neglect any stretching. If this is the case then Eq.~\eqref{eqn:IsotropicTen} can be simplified to~\cite{Seifert1997,Yu2015,Vlahovska2019}
\begin{equation}
	\dfrac{\Delta A}{A_0} = \frac{K_B T}{8\pi\kappa_c}\ln\left(\dfrac{\sigma}{\sigma_0}\right).
\end{equation}

\subsection{Navier-Stokes equations}
Let us consider an initially spherical vesicle with radius $a$ immersed in an externally driven flow. 
The inner and outer fluids are denoted by $\Omega^-$ and $\Omega^+$ while the membrane
separating the two is given by $\Gamma$. 
Assuming both fluids to be Newtonian and incompressible, the flow inside and outside the vesicle can be described by the Navier-Stokes equations:
\begin{equation}\label{eqn:NS}
\rho\frac{D\vec{u^{\pm}}}{Dt}=\nabla\cdot \vec{T}^{\pm}_{hd} \quad \textnormal{with} \quad \nabla\cdot{\vec{u}}^{\pm}=0,
\end{equation}
where $\frac{D}{Dt}$ is the total derivative, $\vec{u}^\pm$ is the velocity field in each fluid, and $\vec{T}_{hd}$ is the bulk hydrodynamic stress tensor given by
\begin{equation}\label{eqn:T_hd}
\vec{T}^{\pm}_{hd} = -p^{\pm}\vec{I}+\mu^{\pm}(\nabla \vec{u}^{\pm} + \nabla^T \vec{u}^{\pm}),
\end{equation}
with the pressure field being $p^\pm$.

The velocity is assumed to be continuous across the interface, $[\vec{u}]=\vec{u}^+-\vec{u}^-=\vec{0}$,
while the hydrodynamic stress undergoes a jump across the interface which is balanced by the interfacial stresses~\cite{vlahovska2007dynamics}
\begin{equation}\label{eqn:StressJump}
	\vec{\tau}_{hd} = \vec{\tau}_{m} + \vec{\tau}_{\gamma} \quad \textnormal{on} \quad \Gamma,
\end{equation} 
where $\vec{\tau}_{hd}=\vec{n}\cdot[\vec{T}_{hd}]$ represents the normal component of hydrodynamic stress with $\vec{n}$ being the outward facing normal vector at the interface.
The bending, $\vec{\tau}_{m}$, and tension, $\vec{\tau}_{\gamma}$, traction forces are given by
\begin{align}
	\vec{\tau}_{m}&=-\kappa_c\big(\frac{H^3}{2}-2KH+\nabla^2_sH\big)\vec{n},\\
	\vec{\tau}_{\gamma}&= \sigma H\vec{n},
\end{align}
where $H$ denotes twice the mean curvature, $K$ is the Gaussian curvature, and $\sigma$ is the isotropic tension, while $\nabla^2_s$ is Laplace-Beltrami
operator. Note that this form assumes zero spontaneous curvature.

Using the Dirac function to take into account the singular contributions of the bending and tension forces, the single-fluid formulation of Navier-Stokes equation can be written as~\cite{Chang1996}
\begin{equation}\label{eqn:ContSurfModel}
	\begin{split}
		\rho\frac{D\vec{u}}{Dt} = & -\nabla p + \nabla\cdot(\mu(\nabla \vec{u} + \nabla^T \vec{u})) \\
								& + \kappa_c \delta(\psi)\bigg(\frac{H^3}{2}-2KH + {\nabla}^2_s H\bigg)\nabla\psi - \delta(\psi)\sigma H \nabla \psi.
	\end{split}
\end{equation}
In this relation, $\psi$ is the level set function defined to describe the evolution of the interface such that $\Gamma$ is given by $\psi=0$
while $\psi<0$ in $\Omega^-$ and $\psi>0$ in $\Omega^+$. The Delta function, $\delta(\psi)$, localizes the singular forces near the interface. Fluid properties can be related
to the level set field via a Heaviside function, $\theta$. For example, viscosity at a location is given by $\mu=\mu^-+\theta(\psi)(\mu^+-\mu^-)$.
More details about this formulation can be found in the literature~\cite{Chang1996,Salac201697,Kolahdouz2015b}. 

\subsection{Non-dimensional model}
In case of a shear or elongational flow, the characteristic time scale corresponding to shape deformation can be defined as 
\begin{equation}
t_{\dot{\gamma}} = 1/\dot{\gamma} \quad \textnormal{or} \quad t_{\dot{\epsilon}} = 1/\dot{\epsilon},
\end{equation}
where $\dot{\gamma}$ is the shear rate and $\dot{\epsilon}$ is the elongation flow strength.
The time scale characterizing the membrane bending rigidity is
\begin{equation}
	t_{\kappa} = \mu^+a^3/\kappa_c.
\end{equation}
The area-dependent tension provides a relaxation time scale given by
\begin{equation}
	t_{\sigma} = \mu^+a/\sigma_{0}.
\end{equation}
For typical values of membrane and fluid properties, $a=10\;\mu$m, $\rho \approx 10^3$ kg/m$^3$, $\mu \approx 10^{-3}$ Pa s, $\kappa_c \approx 10^{-19}$ J, and $\sigma_0 \approx 10^{-7}$ N/m,
the bending time scale is $t_{\kappa}\approx 10$ s, while the tension relaxation time scale is $t_{\sigma} \approx 0.1$ s.
For $\dot{\gamma}=1$ s$^{-1}$ the deformation time scale is $t_{\dot{\gamma}}\approx 1$ s,

Using these time scales and normalizing fluid properties by their values of the outer fluid, the non-dimensional model can be written as
\begin{equation}\label{eqn:NondModel}
\begin{split}
\frac{D\vec{\hat{u}}}{Dt} = & -\hat{\nabla} \hat{p} + \frac{1}{\Re}\hat{\nabla}\cdot(\hat{\mu}(\hat{\nabla} \vec{\hat{u}} + \hat{\nabla}^T \vec{\hat{u}})) \\
					& + \frac{1}{\Cab \Re} \delta(\psi)\bigg(\frac{\hat{H}^3}{2}-2\hat{K}\hat{H} + {\hat{\nabla}}^2_s \hat{H}\bigg)\hat{\nabla} \psi \\
                   & - \frac{1}{\Cat \Re}\delta(\psi)\hat{\sigma}\hat{H} \hat{\nabla} \psi
\end{split}
\end{equation}
with $\Re$ being the Reynolds number given by $\Re = \rho a^2/t_0\mu^+$, $\Cab$ and $\Cat$ being the bending and tension capillary 
numbers expressed as $\Cab = t_{\kappa}/t_0$, $\Cat = t_{\sigma}/t_0$, 
where $t_0$ is chosen to be $t_{\dot{\gamma}}$ or $t_{\dot{\epsilon}}$ for shear and elongational flow, and $\hat{\sigma}=\sigma/\sigma_0$ is the normalized tension. 

\subsection{Numerical methods}
The vesicle surface is modeled using a level-set Jet scheme where the membrane
$\Gamma$ is represented using the zero of a mathematical function $\psi$,~\cite{nave2010,seibold2012}
\begin{equation} 
	\Gamma(\vec{x},t) = \{\vec{x}:\psi(\vec{x},t)=0\}.
\end{equation}
In a given flow-field, the membrane motion is captured
using standard advection schemes. Written in Lagrangian form this is
\begin{equation}
	\dfrac{\mathrm{D} \psi}{\mathrm{D} t}=0,
\end{equation}
which indicates that the level set function behaves as if it was a material
property being advected by the underlying fluid field.

The values of the level set function are only known on the grid points. To compute interface information
away from the grid points, interpolation is required. In a level set jet scheme, all the relevant level set information such as the derivatives of the 
level set function are tracked along with the base level set, allowing for higher order interpolation functions without the need to use wide stencils. 
For example, using a jet which consists of the level set function, $\psi$, and the level set gradient, $\nabla\psi$, 
it is possible to construct a cubic Hermite interpolant using only cell-local information.
For details on Jet level-set methods, readers can refer to the work of Seibold et al.~\cite{seibold2012}.

The fluid field is updated via a semi-implicit, semi-Lagrangian, mass-preserving projection method~\cite{Salac201697}.
First, a tentative velocity field is computed using prior information,
\begin{align}
	\dfrac{\vec{u}^{\ast}-\vec{u}_d^n}{\Delta t}=&-\nabla p^n+\dfrac{1}{\Re}\nabla\cdot\left(\mu\left(\nabla
        \vec{u}^{\ast}+\left(\nabla\hat{\vec{u}}\right)^T\right)\right)\nonumber \\
        & + \frac{1}{\Cab \Re} \delta_\varepsilon(\psi)\bigg(\frac{H^3}{2}-2 K H + \nabla^2_s H\bigg)\nabla \psi \nonumber \\
                   & - \frac{1}{\Cat \Re}\delta_\varepsilon(\psi)\sigma H \nabla \psi,
	\label{eq:tentativeVelocity}
\end{align}
where the material derivative is described using a Lagrangian approach with
$\vec{u}_d^n$ being the departure velocity at time $t^n$ and at the location $\vec{x}_d=\vec{x}-\Delta t\vec{u}^n$.
To aid in numerical stability the Delta function is regularized near the interface, see
Ref.~\cite{TOWERS20086591} for details.

After computing the tentative velocity field, it is projected on to the divergence-free velocity space,
\begin{equation}
	\dfrac{\vec{u}^{n+1}-\vec{u}^{\ast}}{\Delta t}=-\nabla \left(r+(1-\theta_\varepsilon(\psi))r_0\right),
	\label{eq:projection}
\end{equation}
where $r+\left(1-\theta_\varepsilon(\psi))r_0\right)$ is the correction needed for the pressure and $\theta_\varepsilon(\psi)$ is
the regularized Heaviside function~\cite{Towers2009}.
This correction is split into spatially varying, $r$, and spatially constant, $r_0$, parts to enforce 
local ($\nabla\cdot\vec{u}^{n+1}=0$) and global ($\int_\Gamma \vec{u}^{n+1}\cdot\vec{n}\;dA=0$) volume constraints.
After applying the local and global volume constraints to Eq.~\eqref{eq:projection} it is possible
to solve for $r$ and $r_0$.
The pressure is then updated by including the corrections,
\begin{align}
	p^{n+1}&=p^n+r+\left(1-\theta_\varepsilon(\psi)\right)r_0.\label{eq:pressureCorrection}
\end{align}
For full details on the projection algorithm readers are referred to Ref.~\cite{Salac201697}.

\begin{figure}
	\centering
	\begin{subfigure}[t]{8.1cm}
		\includegraphics[width=8.1cm]{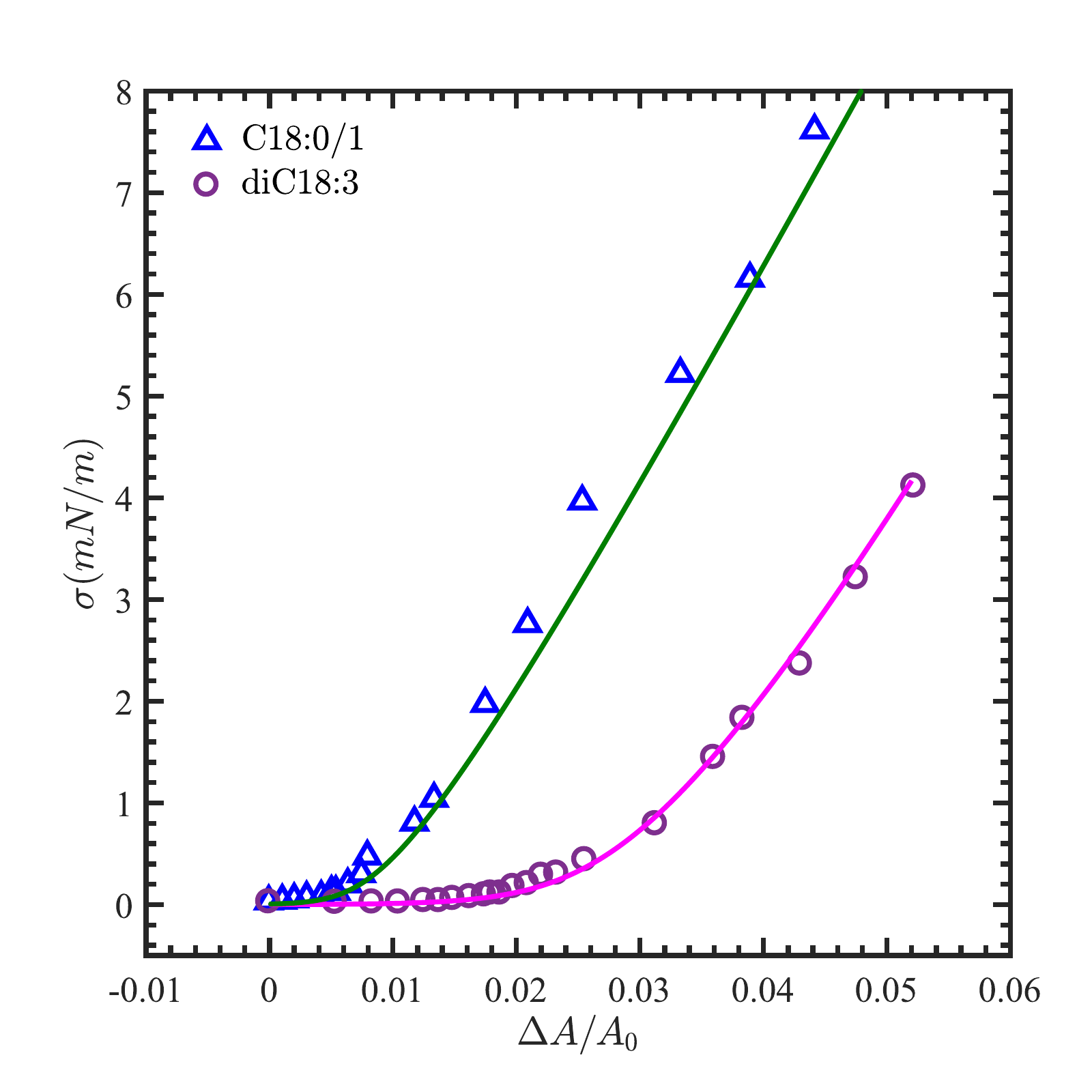}
		\caption{}\label{fgr:T}
	\end{subfigure}
	\begin{subfigure}[t]{8.1cm}
		\includegraphics[width=8.1cm]{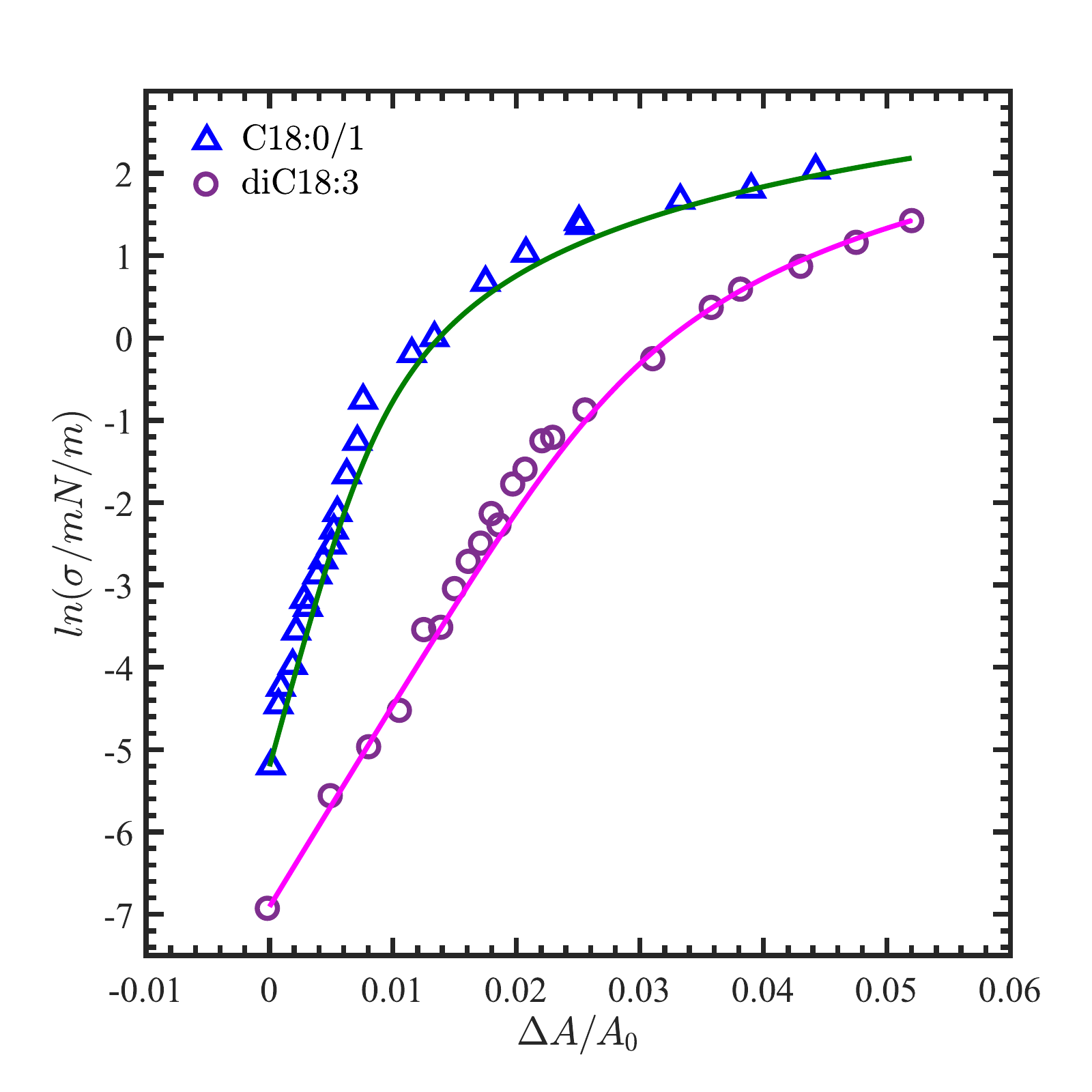}
		\caption{}\label{fgr:logT}
	\end{subfigure}
	\caption{An example of tension as a function of relative area change. Triangles and circles show the measured tension for two vesicles made from C18:0/1 
		and diC18:3 PC~\cite{rawicz2000effect} and the lines show the tension by equation \eqref{eqn:CompleteTen}.}\label{fgr:tension}
\end{figure}

To determine the tentative velocity field, Eq.~\eqref{eq:tentativeVelocity}, it is necessary to determine the membrane tension. The tension is computed based on the current membrane area, which is given by $A=\int_\Omega \delta_\varepsilon(\psi)\;\textnormal{dV}$.
This integral is approximated as a summation over all grid points: $\sum\delta_{i,j,k}dV$
where grid points are given by $\vec{x}_{i,j,k}$, $\delta_{i,j,k}=\delta(\psi(\vec{x}_{i,j,k}))$, $dV$ is the 
volume of each cell surrounding a grid point, and the Dirac function is approximated via the method shown in Towers~\cite{TOWERS20086591}.
According to Eq.~\eqref{eqn:CompleteTen}, $\sigma$ is a nonlinear function of $A$, which needs to be solved numerically at every time step.
In this work the equation is solved via the GSL Library~\cite{GSL}.
Verification of this approach is provided in Fig.~\ref{fgr:tension}, which shows a comparison between the calculated tension and experimentally determined values~\cite{rawicz2000effect} 
for vesicles constructed from C18:0/1 PC ($\kappa_c=0.9\times10^{-19}$ J and $\sigma_0 = 1\times10^{-6}$ N/m)
and diC18:3 PC ($\kappa_c=0.4\times10^{-19}$ J and $\sigma_0 = 5.0\times10^{-6}$ N/m). In both cases $K_a = 230$ mN/m.

During every time step the fluid field is updated based on the interface location from the prior time step. 
The final component is the advancement of the interface due to the flow field. The 
updated fluid field is then used to update the interface location. In particular a semi-implicit and 
semi-Lagrangian scheme is used to advect the level set:
\begin{equation}
	\dfrac{\psi^{n+1}-\psi_d^n}{\Delta t}=0.5\nabla^2\left(\psi^{n+1}-\psi^n\right),
\end{equation}
where $\psi_d^n$ is value of the level set at time $t^n$ and departure location $\vec{x}_d=\vec{x}-\Delta t\vec{u}^{n+1}$.
This method has been shown to increase the stability of numerically stiff moving interface problems~\cite{Velmurugan2016}.

\section{Comparison with Experiments}
In this section, the numerical model described above is compared
to published experimental results for vesicles 
in shear and elongational flow, along with ellipsoidal relaxation.
The goal here is to demonstrate that using physically
realistic material properties, the general model can be used to investigate
a wide variety of flow conditions.
Unless otherwise stated, the inner and outer fluid are assumed to have
the density and viscosity of water at $30^\circ$C,
$\rho^+=\rho^-=995.67$ m$^3$/kg and $\mu^+=\mu^-=7.97\times 10^{-4}$ Pa s,
while the membrane stretching modules is $K_a=0.2$ N/m. In certain cases the inner fluid viscosity
is varied and this is clearly stated in the appropriate sections.
In all cases the physical parameters are used to determine the non-dimensional values, such as Reynolds number,
and the simulations are performed in a domain of physical size $[-4a,4a]^3$ while $N=129$ grid points have been used in each direction.
To better compare to published values, the results are then converted back to physical units where appropriate.

\subsection{Vesicle deformation in shear flow}
Vesicles in shear flow is a classic example of vesicle dynamics and has been widely
studied theoretically~\cite{vlahovska2007dynamics,Misbah2006,Lebedev2007, Farutin2010} and experimentally~\cite{Keller1982,kantsler2005orientation, Kantsler2006}. 
In this work the shear flow is imposed by setting $\vec{u} = (\dot{\gamma}y,0,0)$ on the wall boundaries
in the $y$-direction,
while periodic boundary conditions are applied in $x$- and $z$-directions.

\begin{figure}
	\centering
	\includegraphics[width=5cm]{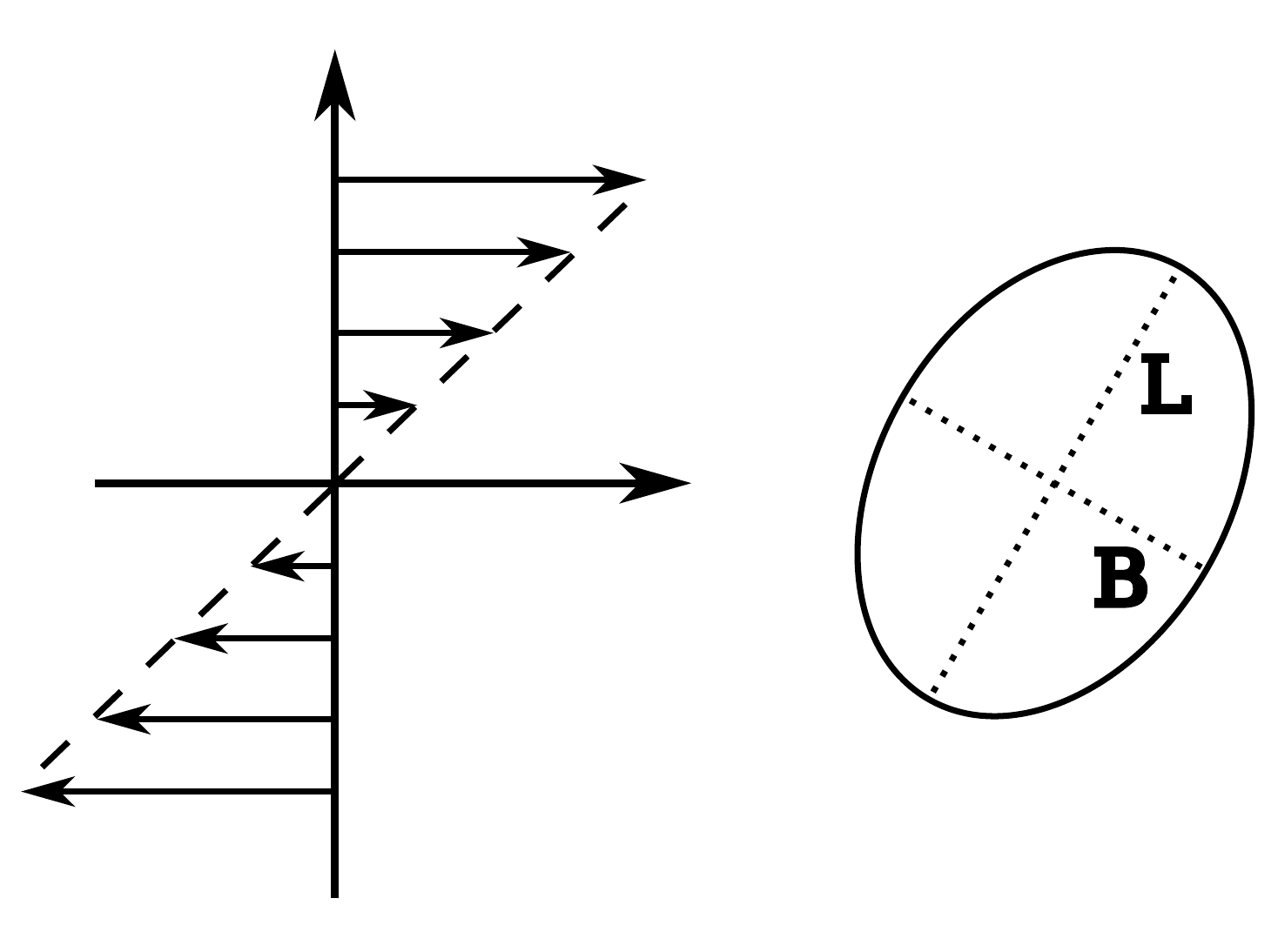}
	\caption{Schematic of a deformed vesicle in a shear flow.}
	\label{fgr:sketch}
\end{figure}

When the viscosity ratio between the inner and outer fluid, $\eta= \mu^-/\mu^+$, is below 
a critical value the vesicle is in the so-called tank-treading regime,
where the vesicle reaches an equilibrium angle with respect to the shear flow~\cite{Kantsler2006, Misbah2006}.
The configuration of a vesicle in such a situation can be measured by two quantities. 
The first is the Taylor deformation parameter~\cite{taylor1934the}, given by
\begin{equation}\label{eqn:TaylorD}
	D = \frac{\mathrm{L-B}}{\mathrm{L+B}}
\end{equation}
where $L$ and $B$ are the lengths of the major and minor axes
of the vesicle, respectively (see Fig.~\ref{fgr:sketch}). Note that in this case the axis lengths
are determined via an ellipsoid with the same inertial tensor~\cite{Ramanujan1998,salac2012reynolds}.

\begin{figure}
	\centering
	\includegraphics[width=8.1cm]{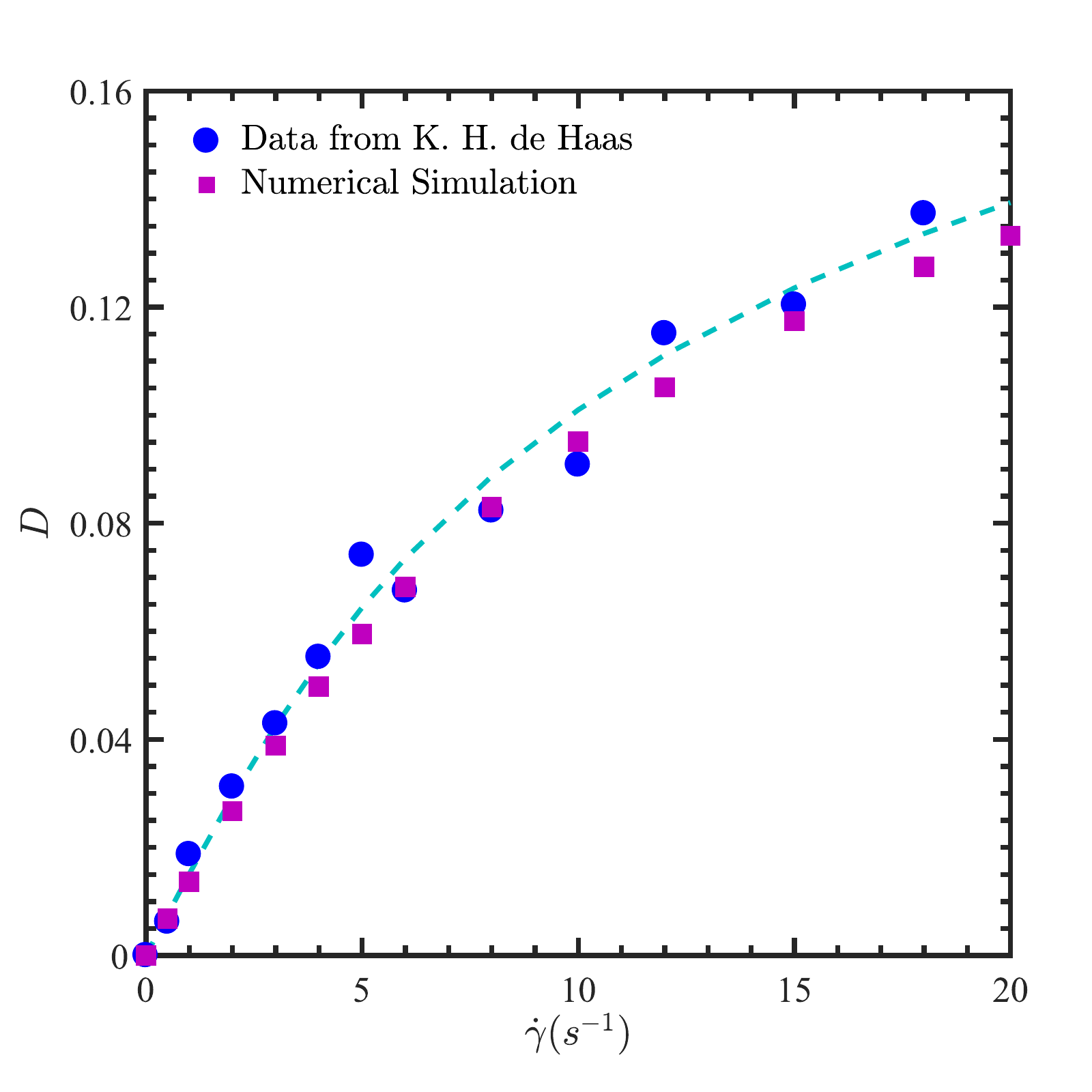}
	\caption{Comparison of vesicle deformation parameter as a function of shear rate from
	Eq.~\eqref{eqn:ShearDef} (dash blue line), the numerical simulation (squares), and experimental results (circles)~\cite{deHaas1997deformation}. 
	The parameters are $a = 55\;\mu\textnormal{m}$, 
		$\kappa_c = 1.3 \times 10^{-20}\;\textnormal{J}$, 
		$\mathrm{\sigma_0} = 3.5 \times 10^{-6}\;\textnormal{N/m}$, and $\mu^+=7.97\times 10^{-4}$ Pa s.}
	\label{fgr:shearDef}
\end{figure}

First consider the equilibrium deformation parameter of an initially spherical vesicle 
under varying shear rates.
The experimental work of de Haas et al. provided a relationship between the applied shear rate
and the equilibrium deformation parameter~\cite{deHaas1997deformation},
\begin{equation}\label{eqn:ShearDef}
	\mathrm{\dot{\gamma}} = \frac{4\mathrm{\sigma_0}D}{5a\mathrm{\mu^+}}\exp\left(\frac{64\mathrm{\pi}\kappa_c}{15K_BT}D^2\right).
\end{equation}
Here it is assumed that the viscosity ratio between the 
inner and outer fluid is $\eta=1$ while the other parameters are those
reported in Ref.~\cite{deHaas1997deformation}. The numerical results and that of de Haas are shown in Fig.~\ref{fgr:shearDef}.
The numerical model is in excellent qualitative agreement with both the experimental and theoretical deformation parameter.
As explained in de Haas, the relatively low bending rigidity obtained could be due to impurities in the vesicle membrane~\cite{deHaas1997deformation}.

\begin{figure}
  \centering
    \begin{subfigure}[t]{1.5cm}
    \includegraphics[width=\linewidth]{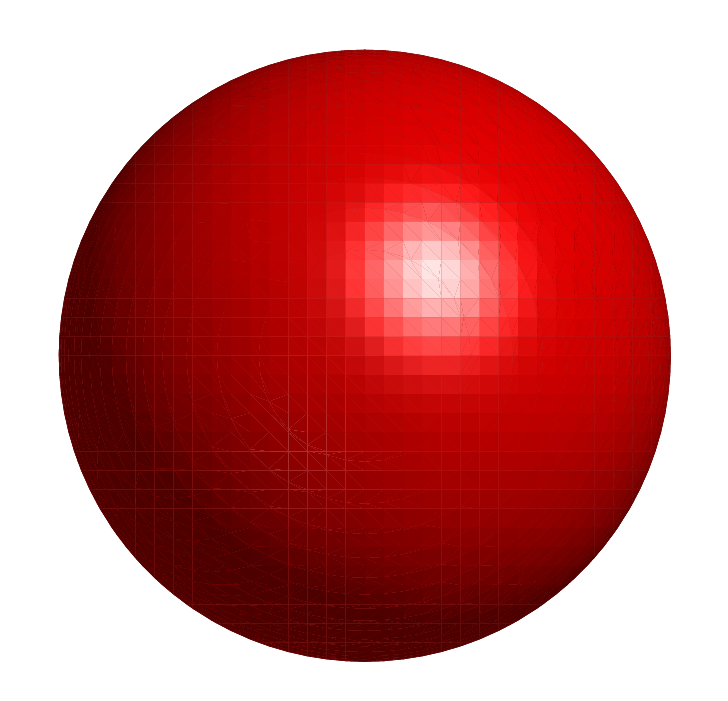}
    \caption{\centering t=0.0}
  \end{subfigure}
  \begin{subfigure}[t]{1.5cm}
    \includegraphics[width=\linewidth]{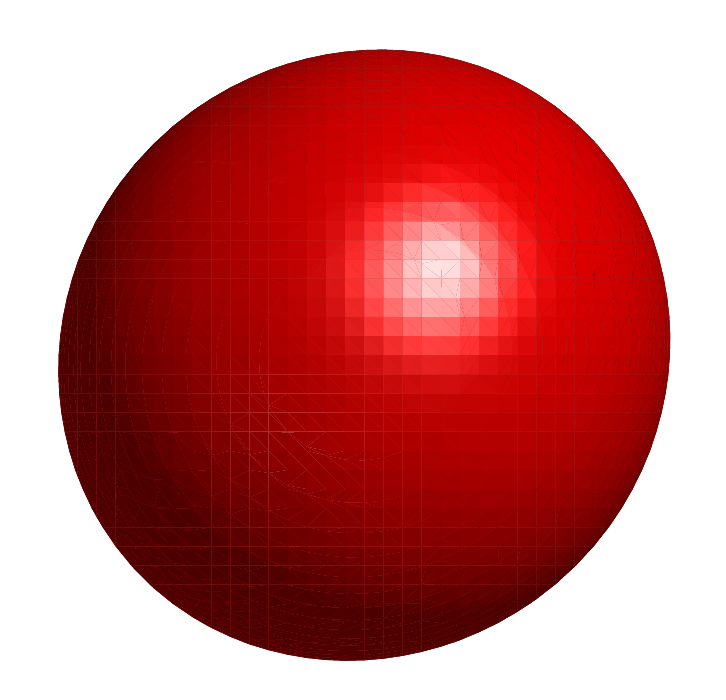}
    \caption{\centering t=0.02}
  \end{subfigure}
   \begin{subfigure}[t]{1.5cm}
    \includegraphics[width=\linewidth]{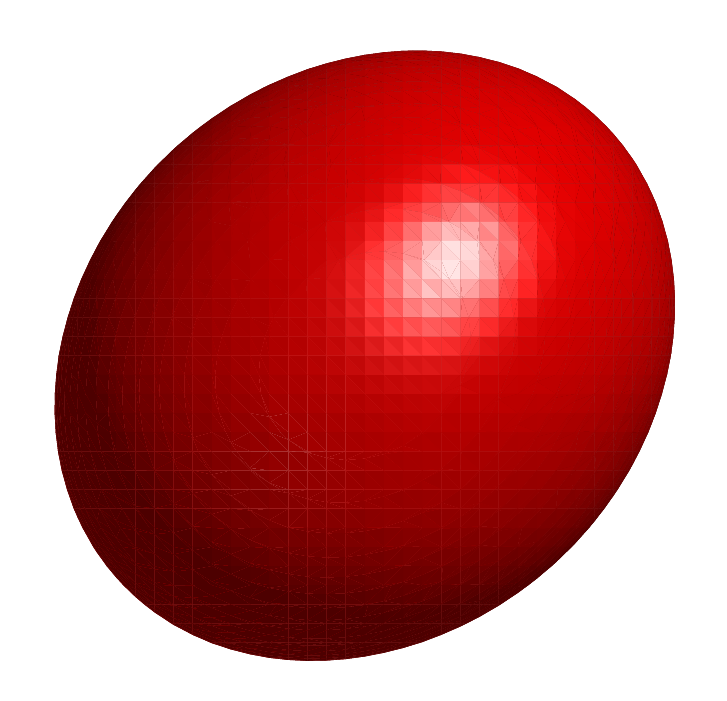}
    \caption{\centering t=0.03}
  \end{subfigure}
  \begin{subfigure}[t]{1.5cm}
    \includegraphics[width=\linewidth]{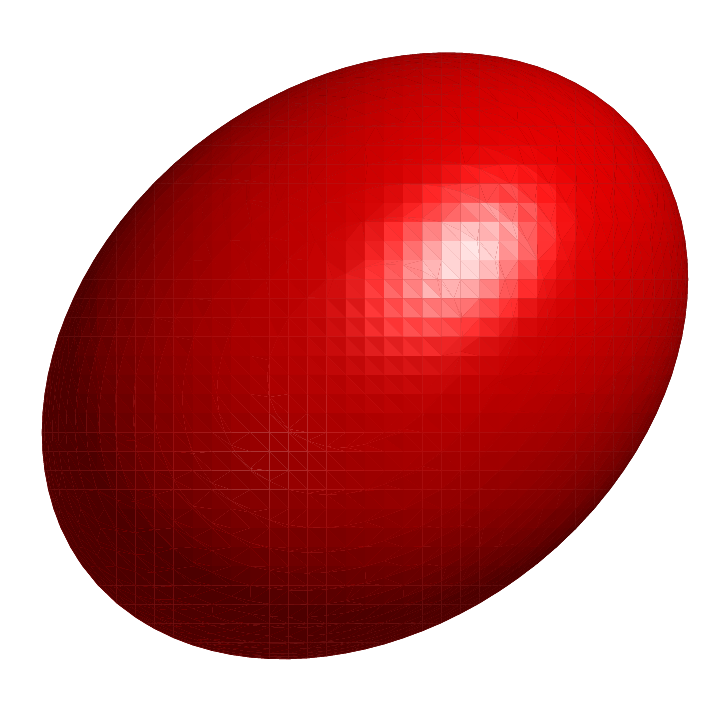}
    \caption{\centering t=0.05}
  \end{subfigure}
   \begin{subfigure}[t]{1.5cm}
    \includegraphics[width=\linewidth]{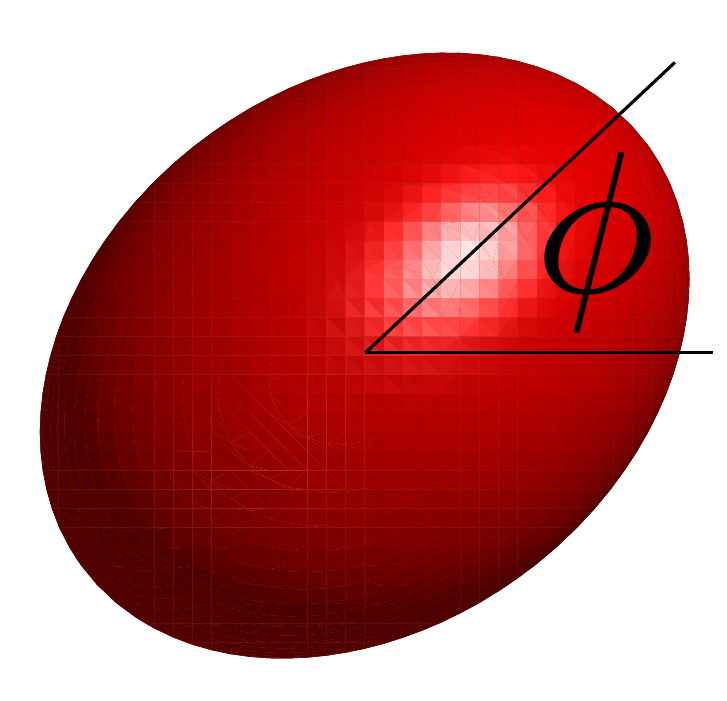}
    \caption{\centering t=0.10}\label{fgr:Axis_Angle}
  \end{subfigure}
 	\caption{Time evolution of a vesicle in shear flow for times $t=0.0,\;0.02,\;0.03,\;0.05,$ and $0.10$ s, from left to right. 
 	The inclination angle of the vesicle can be seen in (e). All the parameters are the same as Fig.~\ref{fgr:shearDef} for $\dot{\gamma}=20$ s$^{-1}$.}
  \label{fgr:TTangle}
\end{figure}

The second common measure of vesicles in shear flow is the equilibrium inclination angle, $\phi$.
The inclination angle of a vesicle is defined as the angle between its major axis and the flow direction, 
as shown in Fig.~\ref{fgr:TTangle}. 
According to linear small deformation theory and in the case of no viscosity contrast, ($\lambda=\eta+1=2$), the stationary inclination angle is related to excess area through~\cite{vlahovska2007dynamics}
\begin{equation}\label{eqn:Phi0}
	\phi_0 = \frac{\pi}{4}-\frac{(9+23\lambda)\Delta^{1/2}}{16\sqrt{30\pi}}.
\end{equation}

\begin{figure}
	\centering
	\begin{subfigure}[t]{8.1cm}
	   \includegraphics[width=1\linewidth]{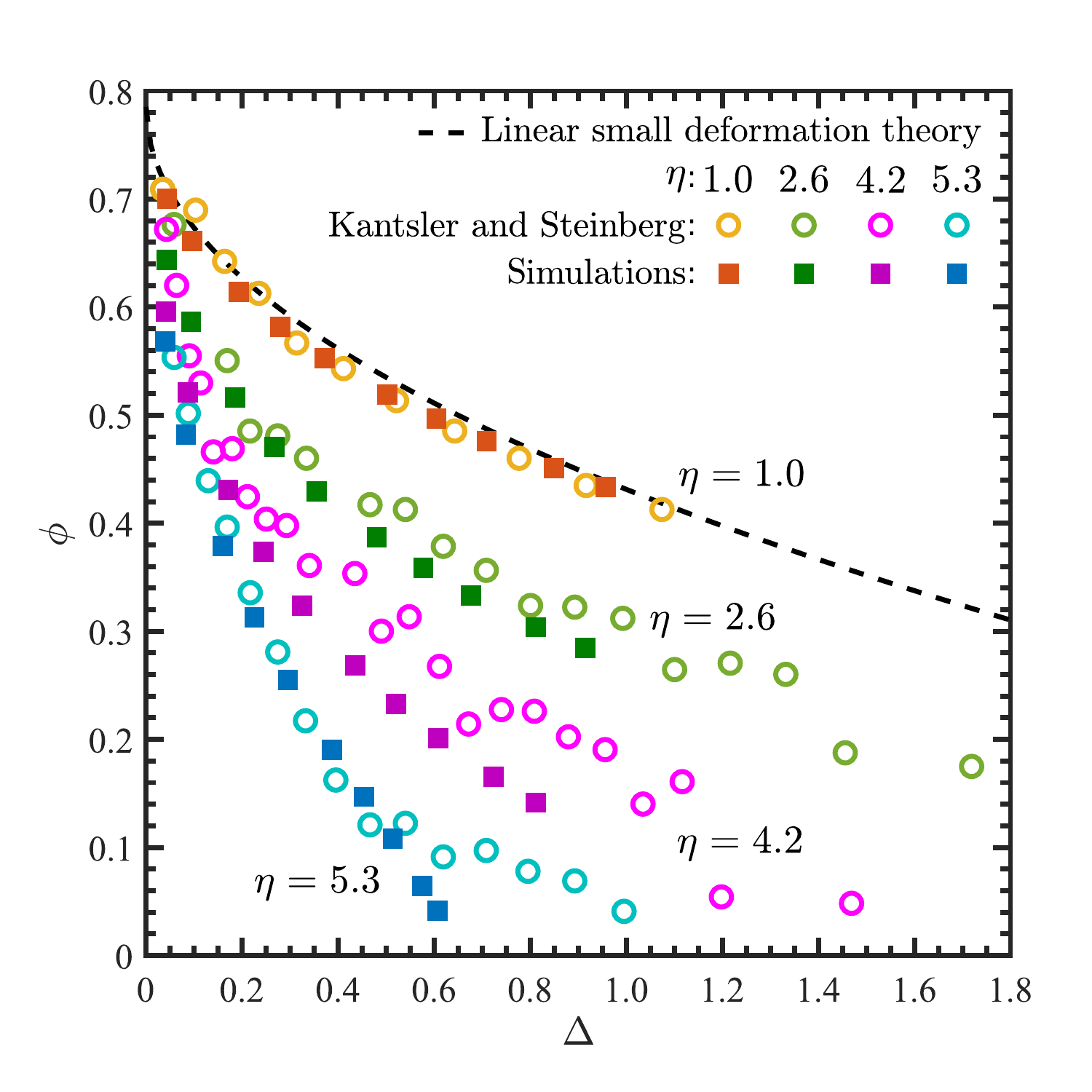}
	   \caption{Inclination angle as a function of excess area.}
	   \label{fgr:KS} 
	\end{subfigure}
	\begin{subfigure}[t]{8.1cm}
	   \includegraphics[width=1\linewidth]{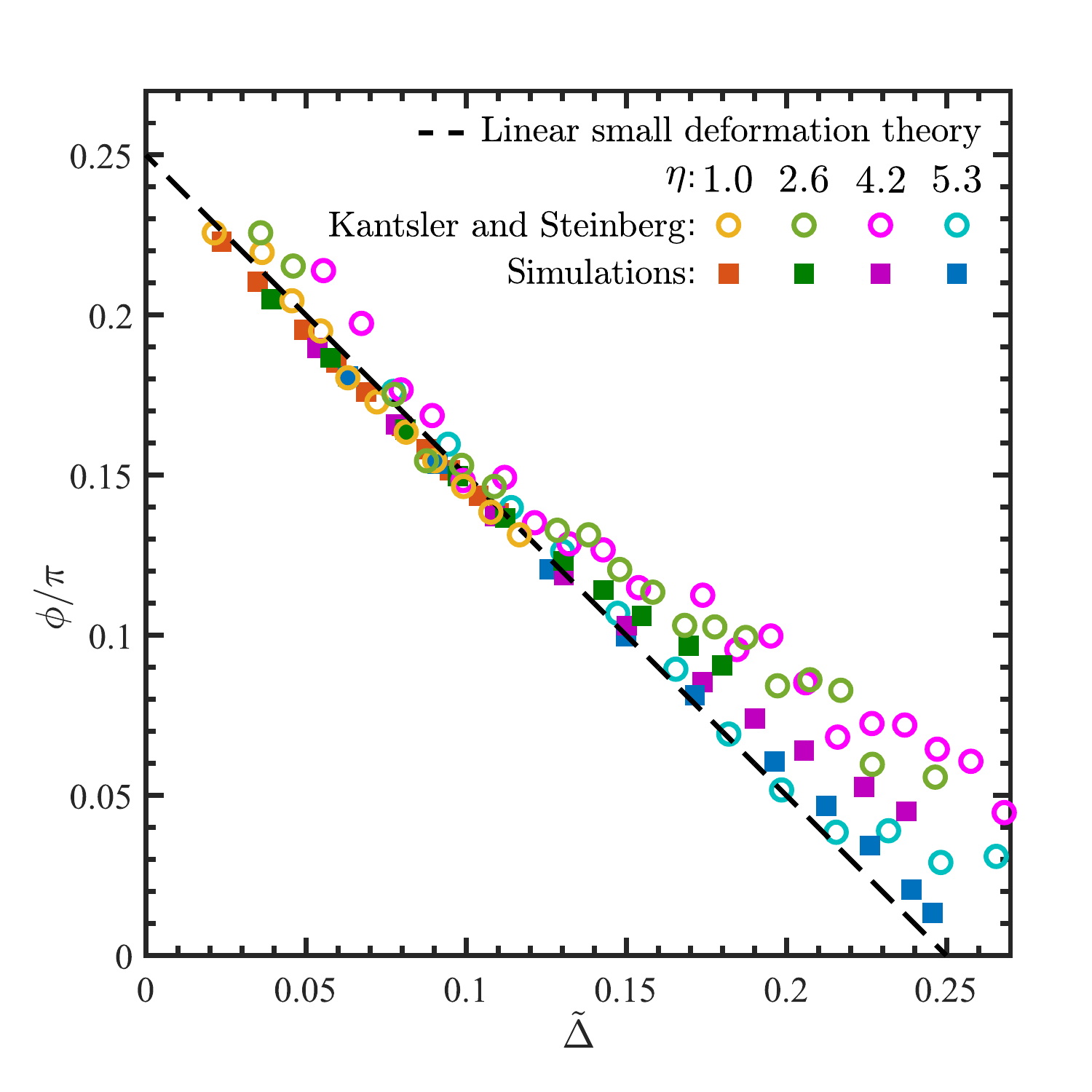}
	   \caption{Inclination angle as a function of re-scaled excess area.}
	   \label{fgr:VG}
	\end{subfigure}
	
	\caption{Stationary inclination angle in radians of vesicles in shear flow as a function of excess area for 
		different viscosity ratios. (a) $\phi$ vs. $\Delta$, for several values of $\eta$, (b) $\phi/\pi$ as a function of rescaled excess 
		area, $\tilde{\Delta}$, given by Eq~\eqref{eqn:ScaledD} for several values of $\eta$. Circles are experimental data from Ref.~\cite{Kantsler2006}, 
		the dashed line represents the linear small deformation theory from Ref.~\cite{vlahovska2007dynamics} given by Eq.~\eqref{eqn:Phi0},
		and the solid squares are numerical results.}\label{fgr:KSVG}
\end{figure}

Consider the same parameters used in Fig.~\ref{fgr:shearDef} except for the initial tension and viscosity ratio.
To allow for various final excess areas the initial tension is varied between $10^{-8}$ N/m to $10^{-5}$ N/m
while the viscosity ratios is set to either $\eta=1,\;2.6,\;4.2,\;\textnormal{or } 5.3$ to match the experimental results.
The vesicle is then allowed to reach a equilibrium, at which point
the visible excess area and inclination angle are calculated.
The numerical results are compared to the linear small deformation theory of Vlahovska and Gracia~\cite{vlahovska2007dynamics} and the experimental results of Kantsler and Steinberg~\cite{Kantsler2006}, see Fig.~\ref{fgr:KSVG} for both the excess area and a rescaled excess area parameter:
\begin{equation}\label{eqn:ScaledD}
	\tilde{\Delta} = \Delta^{1/2}(9+23\lambda)/(16\pi^{3/2}\sqrt{30}).
\end{equation}

As the initial tension decreases, the deformation, and thus the excess area, increases. As is well known, as the 
excess area of a vesicle increases the inclination angle of the vesicle decreases for a given flow~\cite{Zabusky2011}.
We obtain good qualitative agreement with experimental results, with the highest deviation
occurring when $\eta=4.2$. The cause for this discrepancy is unknown, but Kantsler and Steinberg do note a large scatter in the experimental
results for larger viscosity ratios and small inclination angles~\cite{Kantsler2006}. It might also be possible that tension form used, Eq.~\eqref{eqn:CompleteTen}, has larger errors for higher viscosity ratios and excess areas compared to the matched case.

\begin{figure}
	\centering
	\begin{subfigure}[t]{8.1cm}
		\includegraphics[width=\linewidth]{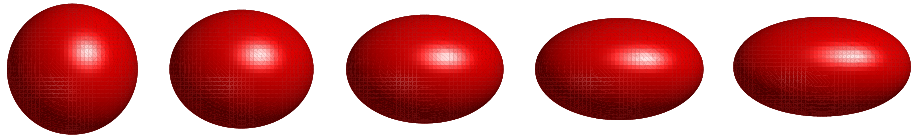}		
		\caption{Sample shapes for $\dot{\epsilon}=0.9$ s$^{-1}$.}\label{fgr:ElongSnapshot}
	\end{subfigure}\\
	\begin{subfigure}[t]{8.1cm}
		\includegraphics[width=\linewidth]{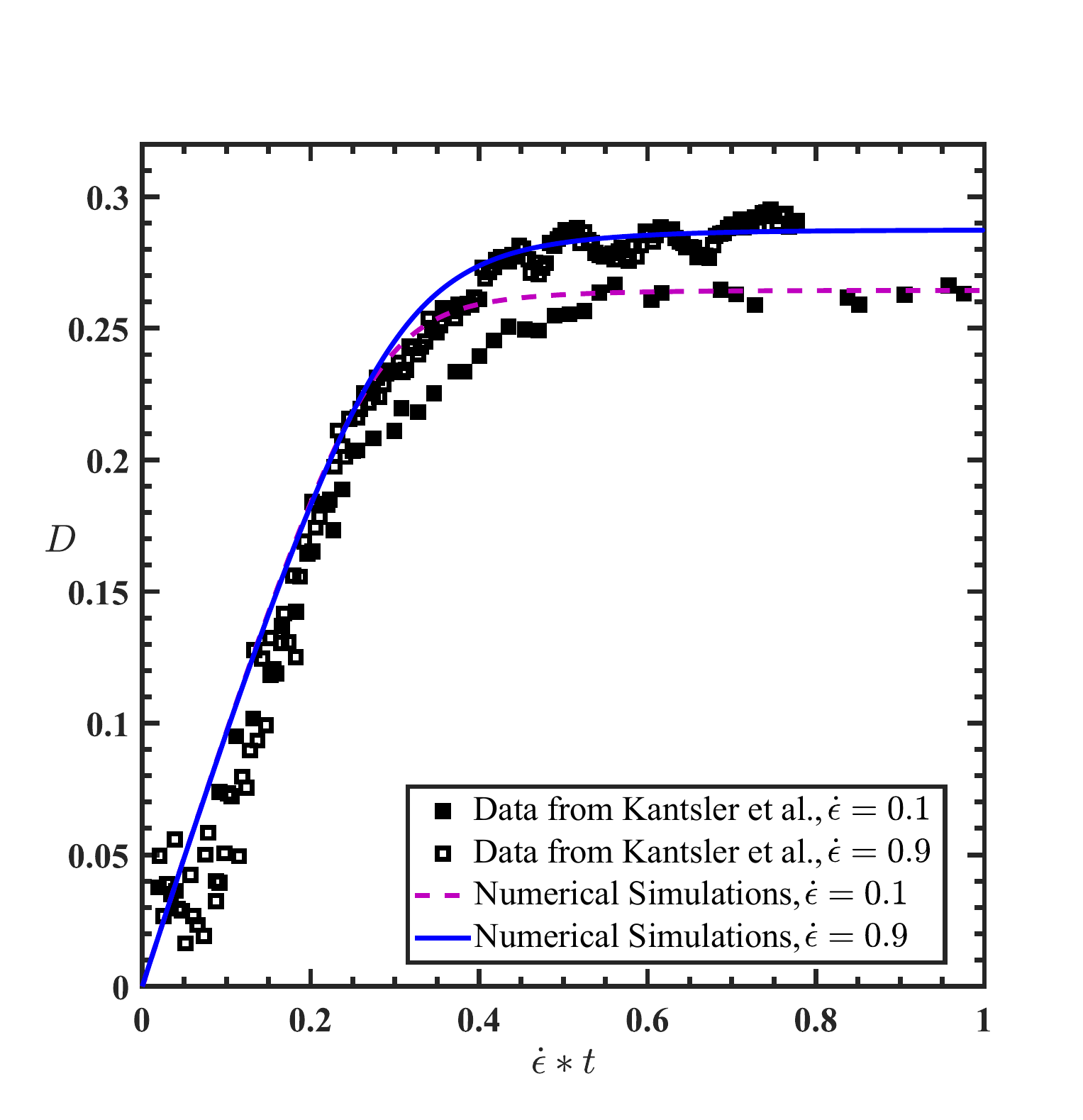}  	
		\caption{Deformation parameter vs. non-dimensional time.}\label{fgr:ElongDef}
	\end{subfigure}
	\caption{Vesicle deformation in elongation flow and comparison with experiments. 
		The vesicle parameters are $\dot{\epsilon}=0.1$ s$^{-1}$: $a = 11.25\;\mu\textnormal{m}$, $\kappa_c = 2 K_BT$, and $\sigma_0= 1.0 \times 10^{-8}\;\textnormal{N/m}$
		while for
		$\dot{\epsilon}=0.9$ s$^{-1}$: $a = 6.25\;\mu\textnormal{m}$, $\kappa_c = 3 K_BT$, and $\sigma_0 = 1.0 \times 10^{-8}\;\textnormal{N/m}$. 
		(a) Deformation of the vesicle with $\dot{\epsilon}=0.9$ s$^{-1}$ for times $t = 0.0,\; 0.1,\; 0.2,\; 0.3,\; 0.5$. (b) Open and filled squares are experiments by Kantsler et al.~\cite{kantsler2007vesicle}, for vesicles with
		$\dot{\epsilon}=0.1$  s$^{-1}$ and $\dot{\epsilon}=0.9$ s$^{-1}$, respectively.
	}
	\label{fgr:Elong}
\end{figure}

\subsection{Vesicle Dynamics in Elongation flow}
We next consider a vesicle in simple elongational flow. A planar flow is 
applied by setting $\vec{u} = (\dot{\epsilon}x,-\dot{\epsilon}y,0)$
on all wall boundaries.
Following Kantsler et al.~\cite{kantsler2007vesicle}, two sample vesicles with matched viscosity are
considered with flow strengths of $\dot{\epsilon}=0.1$ s$^{-1}$ and $\dot{\epsilon}=0.9$ s$^{-1}$.
See Fig.~\ref{fgr:Elong} for sample results and a direct comparison 
of the deformation parameter between the numerical model and 
published experimental results. As with the shear flow examples in the 
prior section the model shows good qualitative agreement with 
the experimental results. We do have a mismatch for the $\epsilon=0.1$ case in the transition from linear growth to the equilibrium deformation
around a normalized time of 0.3. 
It is suspected that the ratio between the numerical dissipation from the underlying discritizations and the applied flow is the cause,
as this difference is not seen for the $\epsilon=0.9$ case. Despite this, the equilibrium deformation parameter for both cases matches
well with the experimental result.

\subsection{Relaxation of deformed vesicle}

The final comparison will consider the relaxation of initially elliptical vesicles back to a spherical shape.
Specifically, the results will be compared to those presented by Yu et al., which used electric pulses to deform an initially spherical vesicle
into an elliptical shape~\cite{Yu2015}. In this work we use the elongational flow described above to deform the vesicle until 
a shape factor $\epsilon=L/B-1=0.15$, where $L$ and $B$ are the long and short axis of the vesicle, is achieved. At that
point the velocity is set to zero on the boundary and the shape factor is observed as a function of time.
Despite the mechanism which induces the elongation differing from Yu et al, it will be demonstrated that the relaxation dynamics match,
which confirms the statements in Yu et al.

\begin{table}[H]
  \begin{center}
    \caption{Properties of 8 POPC vesicles in Fig.~\ref{fgr:eps}. Values are those
		given in supplementary material for Ref.~\cite{Yu2015}.}
    \label{tab:POPCprop}
    \begin{tabular}{|c|c|c|c|} 
     \hline 
      \textbf{Parameter} & $a$ ($\mu$m) & $\kappa_c$ ($10^{-19}$J) & $\sigma_0$ ($\mu$N/m) \\
      \hline 
      \hline
      \textbf{Vesicle $\#1$} & 21.1 & 1.10 & 2.07 \\      \hline
      \textbf{Vesicle $\#2$} & 34.8 & 1.20 & 2.78 \\      \hline
      \textbf{Vesicle $\#3$} & 27.1 & 1.01 & 4.64 \\      \hline
      \textbf{Vesicle $\#4$} & 16.4 & 0.83 & 3.06 \\      \hline
      \textbf{Vesicle $\#5$} & 22.2 & 1.38 & 3.46 \\      \hline
      \textbf{Vesicle $\#6$} & 33.0 & 1.25 & 3.42 \\      \hline
      \textbf{Vesicle $\#7$} & 37.0 & 1.46 & 3.55 \\      \hline
      \textbf{Vesicle $\#8$} & 20.1 & 0.91 & 8.69 \\      \hline
    \end{tabular}
  \end{center}  
\end{table}

\begin{figure}
	\centering
	\includegraphics[width=8.1cm]{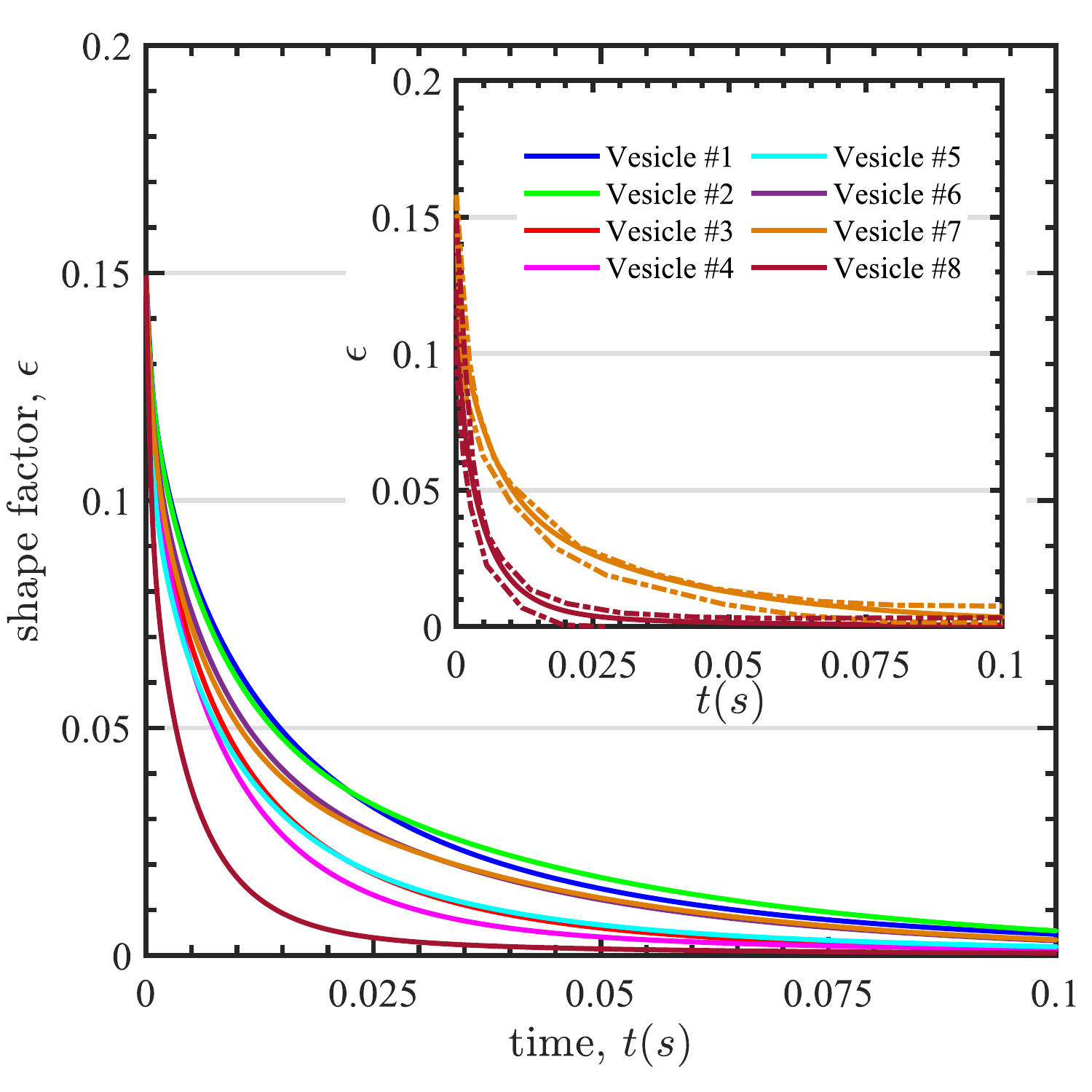}	
	\caption{Shape factor, $\epsilon$, versus time for 8 POPC vesicles with properties given in Table~\ref{tab:POPCprop}. 
		Inset: Comparison with results for Vesicles \#7 and \#8 from the supplementary material of Ref.~\cite{Yu2015}. The dash-dot lines 
		indicate the maximum and minimum deformation parameter observed experimentally.}
	\label{fgr:eps}
\end{figure}

In particular we re-create the relaxation of the 8 POPC vesicles shown in Ref.~\cite{Yu2015} and the associated supplementary materials, see Fig.~\ref{fgr:eps}.
It is assumed that the inner and outer fluid viscosities are $\mu^-=1.253\times 10^{-3}$ Pa s and $\mu^+=1.019\times 10^{-3}$ Pa s
while the vesicle radius, bending rigidity, and initial tension are those provided in Table S2 of the supplementary materials of Yu et al. and are 
provided in Table~\ref{tab:POPCprop} for convenience.
Using the given membrane properties, the numerical model
is able to qualitatively capture the experimental results very well.
This is further confirmed by the inset of Fig.~\ref{fgr:eps},
which includes the results for vesicles \#7 and \#8
which are compared to the shape factor bounds estimated from the supplementary materials of Ref.~\cite{Yu2015}.

\section{Material Property Influence}

Finally, consider the influence of material properties on initially-spherical
vesicles in shear flow. The goal here is to provide some fundamental information on how 
slight variations in quantities such as viscosity ratio can influence measurable quantities. 
This knowledge is particularly important when designing appropriate experimental systems
which could be used to obtain material properties.

\begin{figure}
	\centering
	\begin{subfigure}[t]{8.1cm}
		\centering
		\includegraphics[width=\linewidth]{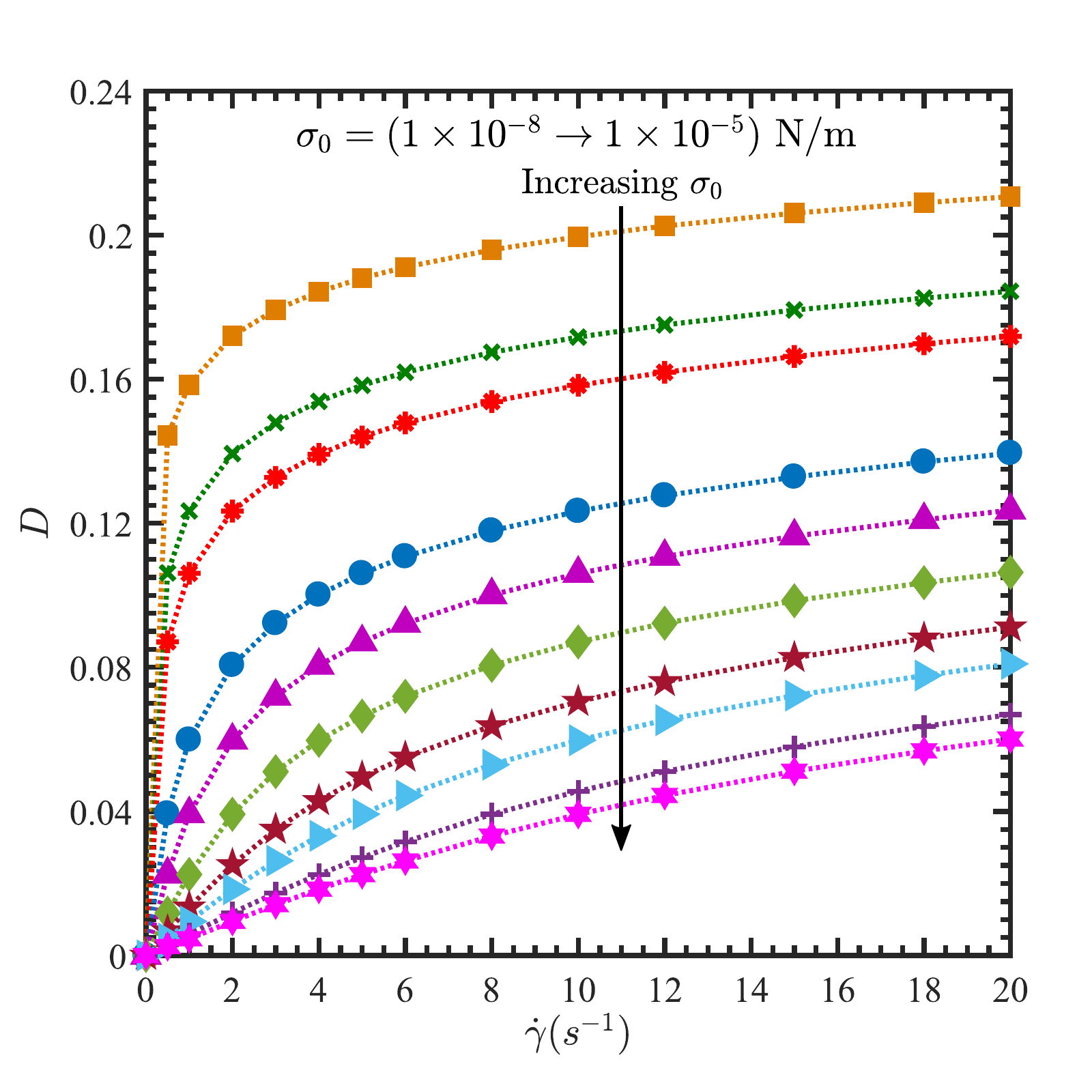}
		\caption{Effect of initial tension on equilibrium deformation parameter.}
		\label{fgr:sigma} 	
	\end{subfigure}
	\begin{subfigure}[t]{8.1cm}
		\centering
	   \includegraphics[width=\linewidth]{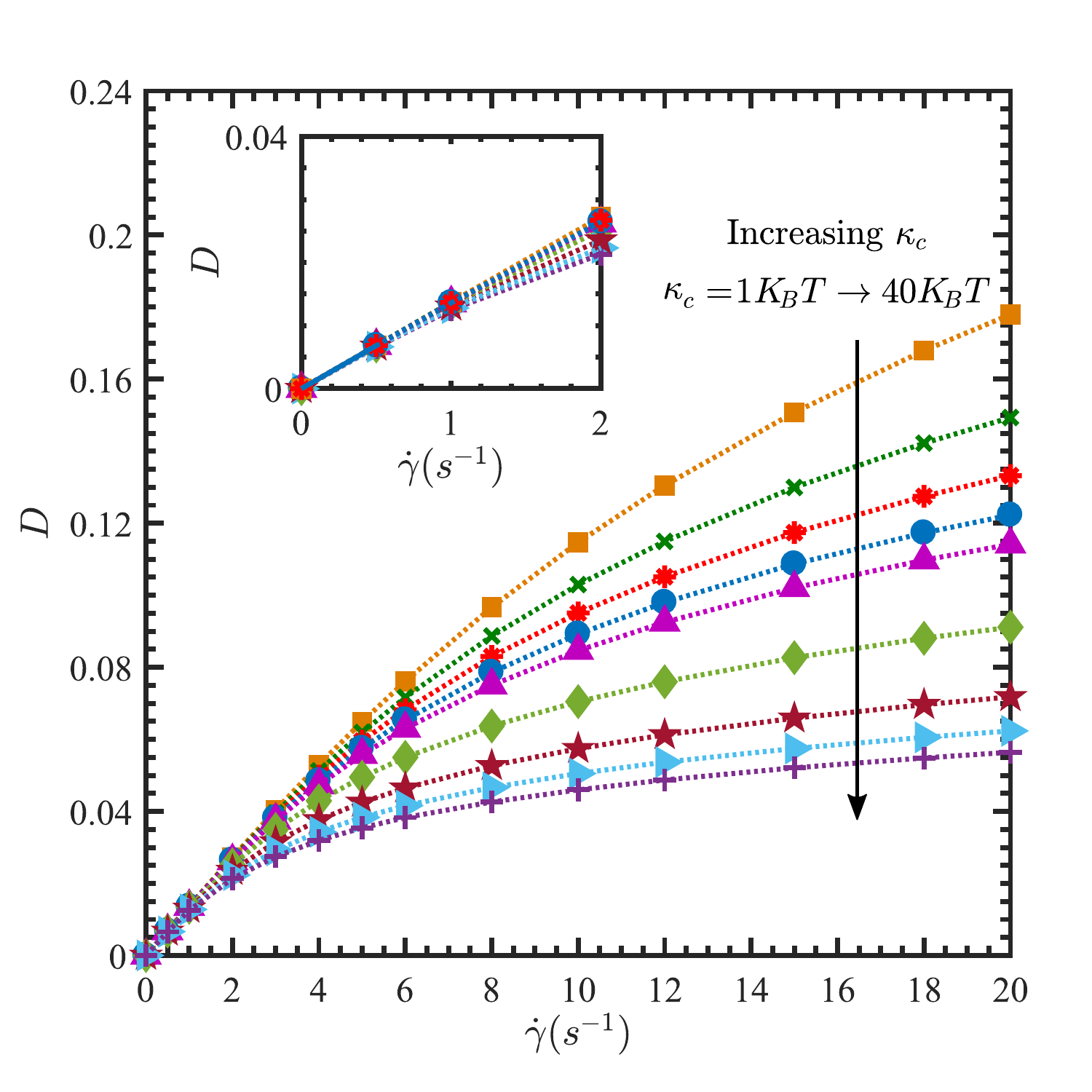}
	   \caption{Effect of bending rigidity on equilibrium deformation parameter.}
	   \label{fgr:kc}
	\end{subfigure}
	\caption{Effect of (a) initial tension and (b) bending rigidity on deformation of a vesicle in shear flow. 
		Initial tension values in (a) are $\sigma_0 = (1.0 \times 10^{-8},\; 5.0 \times 10^{-8},\; 1.0 \times 10^{-7},\; 5.0 \times 10^{-7},\; 1.0 \times 10^{-6},\; 2.0 \times 10^{-6},\; 
			3.5 \times 10^{-6},\; 5.0 \times 10^{-6},\; 8.0 \times 10^{-6},\; 1.0 \times 10^{-5})$ N/m
			and $\kappa_c = 4.33\times10^{-20}$ J. 
			Bending rigidity values in (b) are $\kappa_c = (4.33\times10^{-21},\; 8.66\times10^{-21},\; 1.30\times10^{-20},\; 
			1.73\times10^{-20},\; 2.17\times10^{-20},\; 4.33\times10^{-20},\; 8.66\times10^{-20},\; 
			1.30\times10^{-19},\; 1.73\times10^{-19})$ J and $\sigma_0 = 3.5 \times 10^{-6}$ N/m. 
			The remaining parameters are the same as Fig.~\ref{fgr:shearDef}. 
			The inset in (b) shows a view of the deformation parameter for $\dot{\gamma}=(0\rightarrow2)\;\textnormal{s}^{-1}$ for different bending rigidity values.}\label{fgr:sigmaBending}
\end{figure}

Consider the influence of initial tension and bending rigidity on
the equilibrium deformation parameter for a vesicle in shear flow, Fig.~\ref{fgr:sigmaBending}.
As expected, increasing the initial tension or bending rigidity results in a smaller deformation parameter
at a given shear rate. Of interest is the insensitivity of the deformation parameter to the bending 
rigidity in low-shear conditions, Fig.~\ref{fgr:kc}. This is in contrast
to the high sensitivity of the deformation parameter on the initial tension in this low-shear regime.
This offers the possibility of using low-shear experiments to determine the initial tension and then 
high-shear experiments to determine the bending rigidity (once the initial tension is determined).

\begin{figure}
	\centering
	\includegraphics[width=8.1cm]{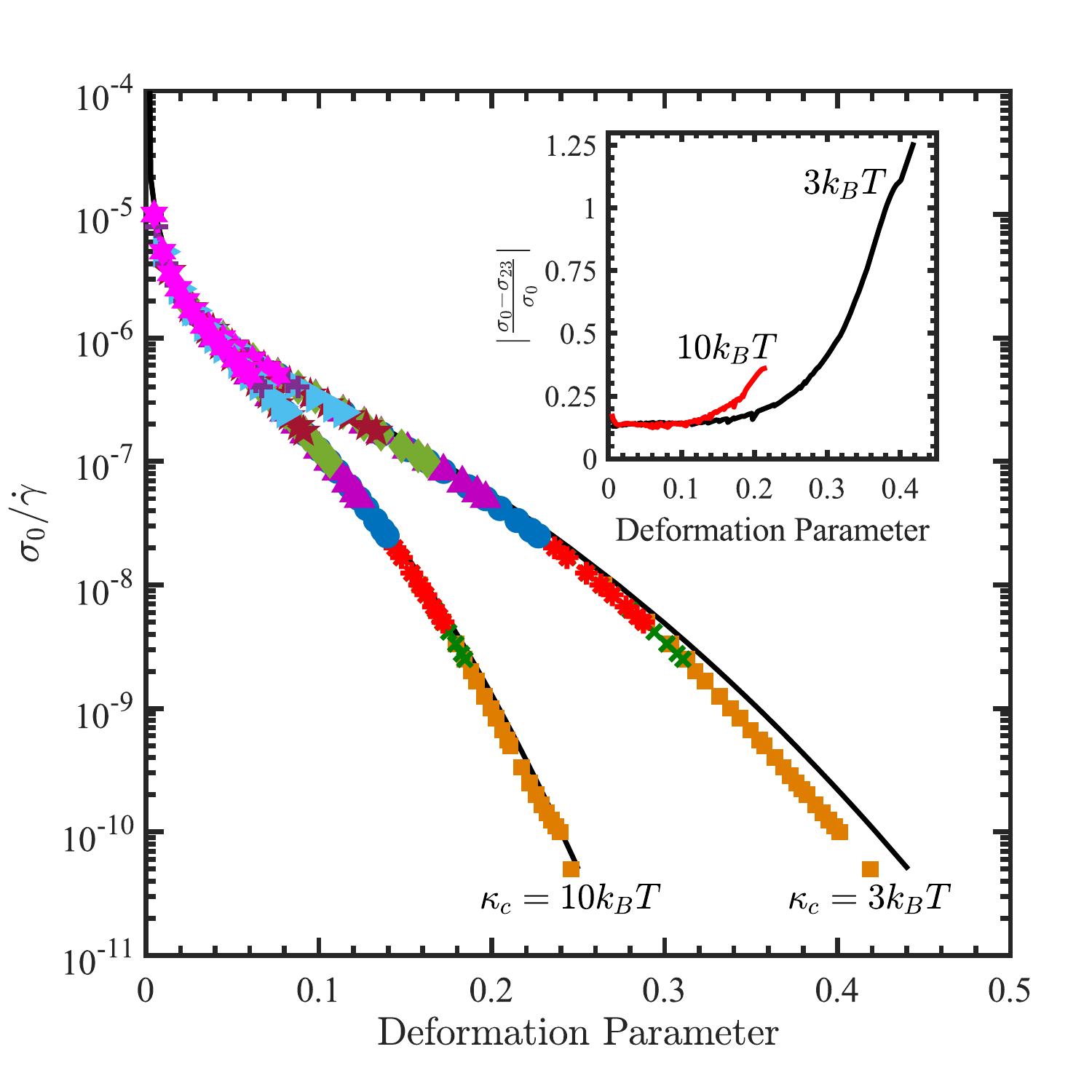}
	\caption{Scaled initial tension vs deformation of a vesicle in shear flow. Varying 
	equilibrium deformation parameters are obtained by varying $\sigma_0$ and shear flow rates.
		The solid lines correspond to Eq.~\eqref{eqn:ShearDef}. Inset: The error between the 
		initial tension calculated from Eq.~\eqref{eqn:ShearDef}, $\sigma_{23}$, and the value used in the simulation, $\sigma_0$,
		for a given deformation parameter.}\label{fgr:sigma0Shear_D}
\end{figure}

A major difficulty in focusing on the low-shear regime is that the resulting deformation parameters
are small and could be subject to large measurement errors. Therefore, it would be advantageous to 
use larger shear rates which results in higher deformation parameters. While this would reduce the
amount of experimental error (compared to the parameters to be measured), this does result in a system
where both the bending rigidity and initial tension play a strong role. Additionally, as 
the deformation parameter increases, closed-form expressions relating the material parameters to the deformation
parameter, such as Eq.~\eqref{eqn:ShearDef}, become less valid. See Fig.~\ref{fgr:sigma0Shear_D} for an example
using two bending rigidities. Differences between the closed-form expression in Eq.~\eqref{eqn:ShearDef} 
and those from the numerical simulation are greater than 25\% at moderate deformations ($D\approx 0.2$) and 
grow very large as $D$ increases. Therefore, it would be necessary to 
use models with fewer assumptions, such as that presented in this work.

\begin{figure}
	\centering
	\includegraphics[width=8.1cm]{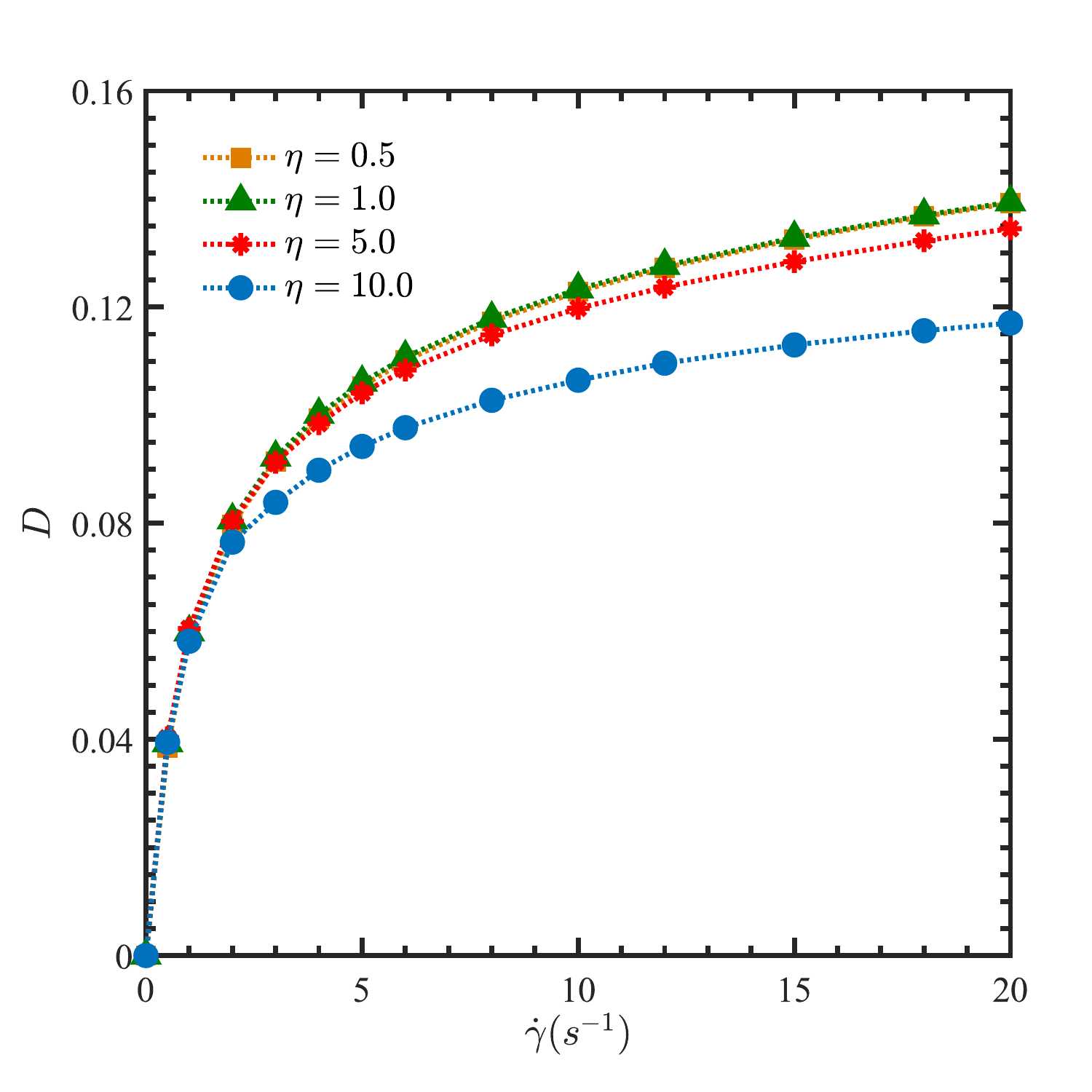}
	\caption{Effect of viscosity ratio on deformation of a vesicle in shear flow. 
		The vesicle parameters are the same as Fig.~\ref{fgr:sigmaBending} and $\kappa_c = 4.33\times10^{-20}$ J, and $\sigma_0 = 5.0 \times 10^{-7}$ N/m.
		}\label{fgr:sigmaViscosity}
\end{figure}

\begin{figure}
	\centering
	\includegraphics[width=8.1cm]{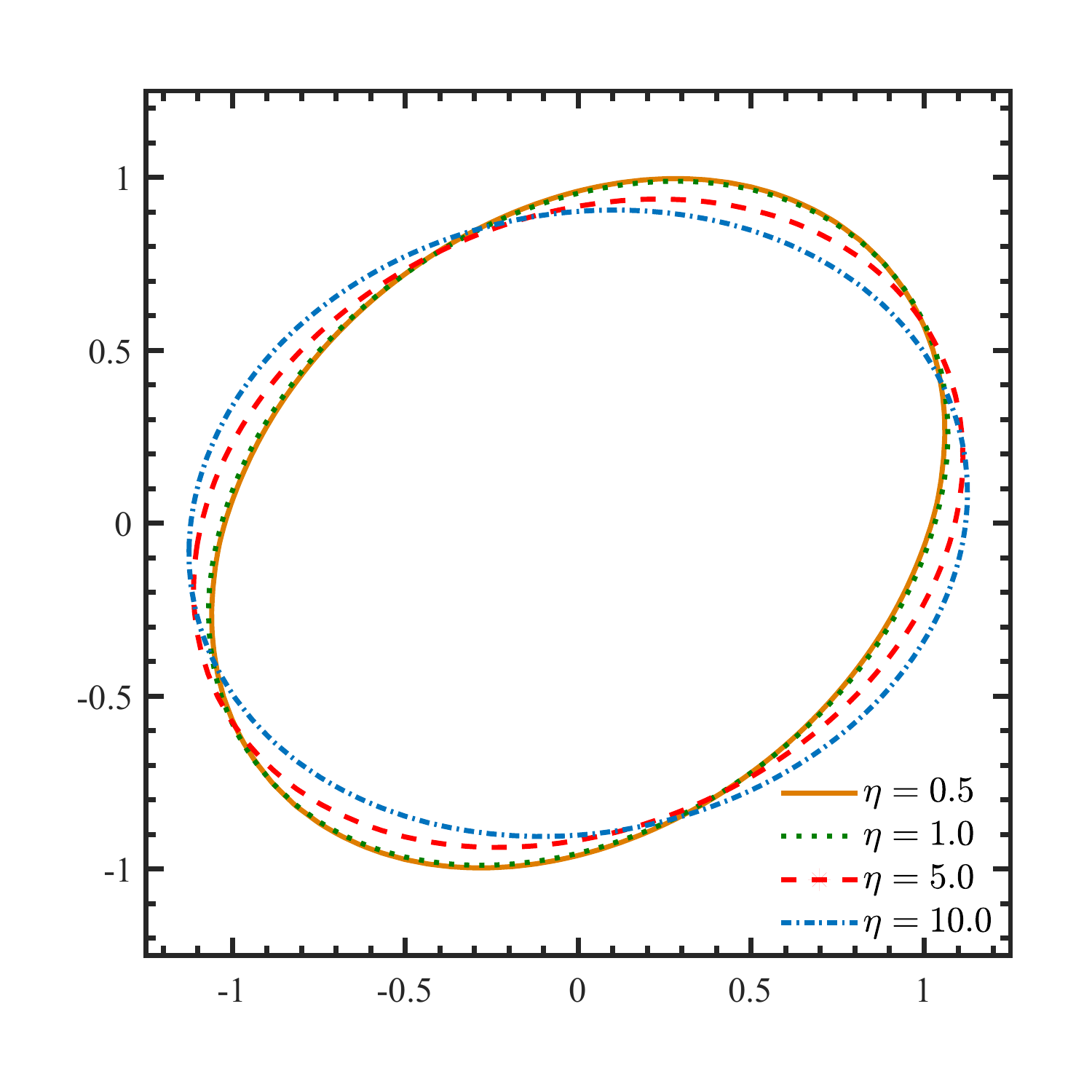}
	\caption{Cross-section of the vesicles in the $x-y$ plane for a shear rate of 20 s$^{-1}$ and four different viscosity ratios, $\eta$. As $\eta$ increases
		the inclination angle of the vesicle decreases, exposing it to smaller shear-induced forces.}\label{fgr:Vis_Shear20_XY}
\end{figure}

\begin{figure}
	\centering	
	\begin{subfigure}[t]{8.1cm}
		\centering
		\includegraphics[width=1\linewidth]{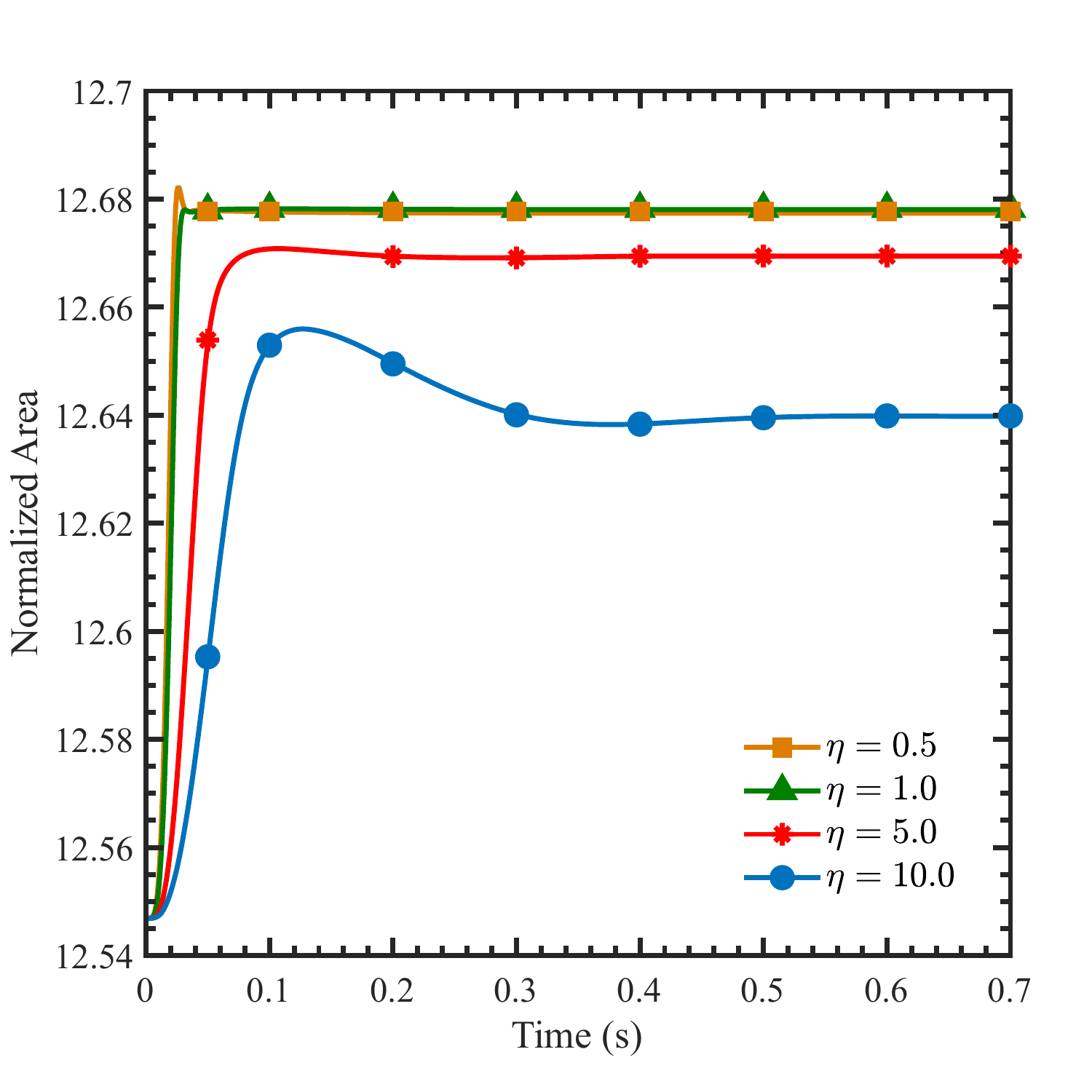}
		\caption{Surface area vs time.}
		\label{fgr:Vis_Shear20_Area} 	
	\end{subfigure}
	\begin{subfigure}[t]{8.1cm}
		\centering
	   \includegraphics[width=1\linewidth]{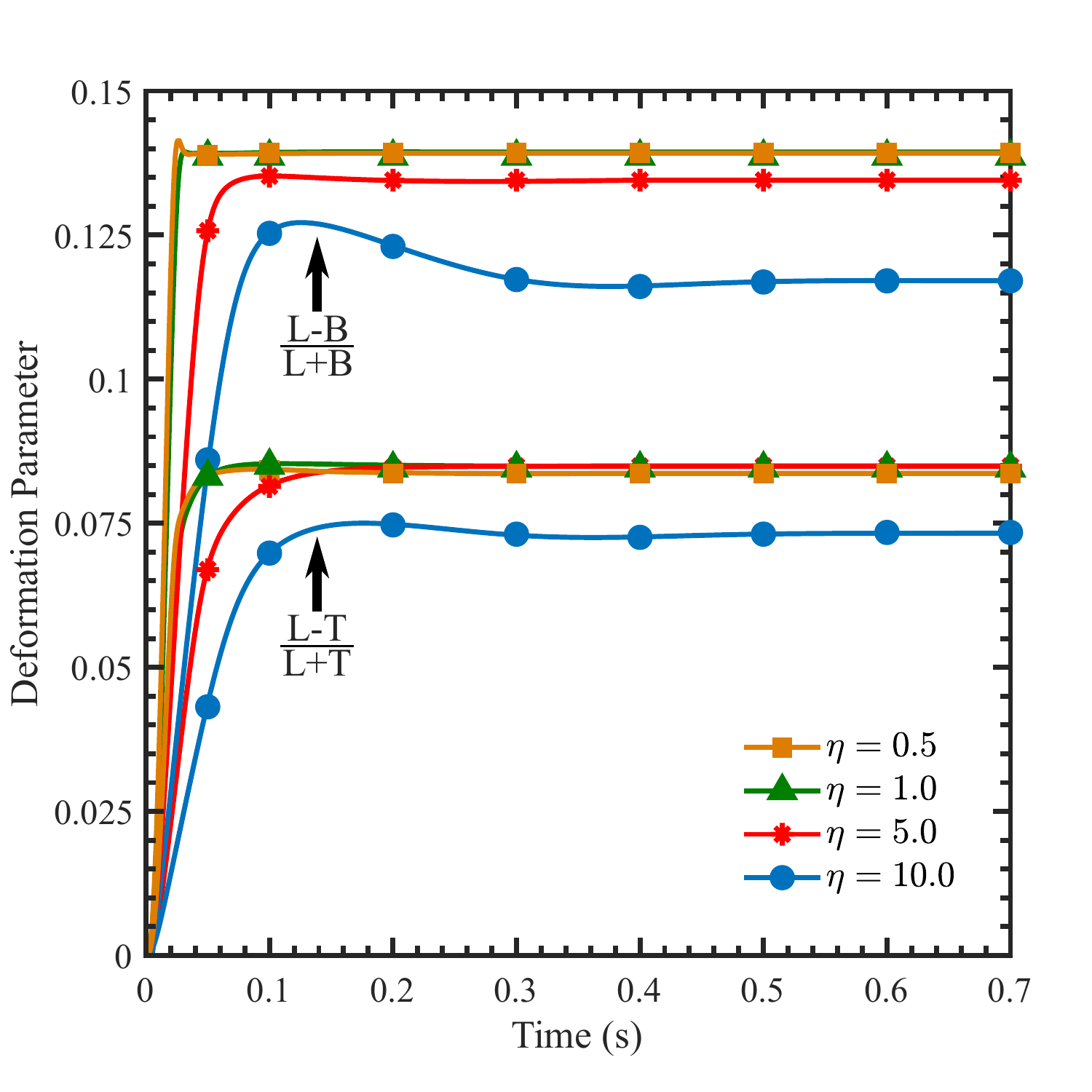}
	   \caption{Deformation parameters vs time.}
	   \label{fgr:Vis_Shear20_D}
	\end{subfigure}	
	
	\caption{Normalized area and deformation parameters of vesicles with varying viscosity ratios over time in a shear flow with a strength of 20 s$^{-1}$.
	The deformation parameters use the long (L), mid (T), and short (B), axis lengths. The mid-axis corresponds the 
		size of the vesicle in the $z$ (out-of-plane) direction in Fig.~\ref{fgr:Vis_Shear20_XY}. Normalized area is given by $A/a^2$, where $A$ is the visible area and $a$ is the radius of the initially spherical vesicle.
	}\label{fgr:Vis_Shear20_Param}
\end{figure}

Finally, consider the effect of viscosity ratio on the equilibrium deformation, Fig.~\ref{fgr:sigmaViscosity}. 
Here, four viscosity ratios are considered: $\eta=0.5,1.0,5.0,\textnormal{ and }10$. The deformation 
parameter at various shear rates is shown in Fig.~\ref{fgr:sigmaViscosity}. It is clear that 
the equilibrium deformation parameter is insensitive to viscosity ratio
at low shear rates or if $\eta\leq1$. The deformation parameter demonstrates a clear dependence on the viscosity ratio when it is greater than one.
This can be understood by considering the effect of viscosity ratio on the inclination angle of the vesicle with respect to the shear flow.
It is well known that as viscosity ratio increases the angle between the vesicle and the shear flow decreases~\cite{Kantsler2006,kantsler2005orientation},
which is the case here, Fig.~\ref{fgr:Vis_Shear20_XY}. As the inclination angle has decreased, any shear-induced forces decrease.
This reduction in forces results in smaller deformations, as seen in Fig.~\ref{fgr:Vis_Shear20_Param}. For the cases of $\eta=0.5$
and $\eta=1.0$ there is little difference in the surface area or deformation parameters over time. As the viscosity
ratio increases the vesicle becomes more inclined, which results in smaller area increases and deformation parameters.
It is therefore important to control (and verify) the viscosity ratio when attempting to use shear flow experiments to 
extract membrane properties.

\section{Discussion and Conclusions}

In this work a general numerical framework to model liposome vesicles is extended to account for the exchange between 
sub-optical area and visible area. By not making assumptions about the shape, 
this framework is capable of modeling vesicles in a wide variety of flow conditions.
The validity of the model has been confirmed by direct comparisons with several available experimental results.

The use of shear-flow experiments have previously been used to determine the initial tension and bending rigidity of vesicles.
Using the framework outlined here the influence of material properties on the resulting deformation parameter is explored, 
and it was shown that deviation from the previously used analytic model occurs for large deformation parameters. 
The influence of another common system parameter, the viscosity ratio between the inner and outer fluid, has also been 
explored. It is important for the inner fluid of the vesicle to be less viscous than the outer fluid, as having
a viscosity ratio above one dramatically changes the equilibrium deformation parameter and may lead to incorrect material 
property estimates.

The development of this framework opens up the possibility of more formalized sensitivity analysis of other system 
parameters and their influence on experimentally measurable quantities. For example, the bulk deformation obtained
in a cross-flow experimental setup may be less sensitive to the fluid viscosity ratio or
may show a larger initial tension independent regime than the shear-flow case. It will also be possible to utilize the
model to determine the material properties via the inverse problem from experiments.

While the model here can be used for a wide variety of flow conditions, it can not be used to investigate all possible experiments. In particular, those situations where there is zero sub-optical area additional numerical components must be added to the penalize local stretching of the membrane. One possibility is to specify the total area possible of a vesicle and to determine a spatially-varying membrane tension to enforce local incompressibility, similar to the author's prior
works~\cite{GERA2018,kolahdouz2015,salac2011,salac2012reynolds}. This would allow for the numerical investigations of an even larger number of flow conditions.

\clearpage


%

\end{document}